\documentclass[aps,prl,preprint,groupedaddress,showpacs]{revtex4-1}
\pdfoutput=1
\usepackage{bm}
\usepackage{graphicx}
\usepackage{mathrsfs}
\usepackage{amssymb}
\usepackage{amsmath}
\usepackage{booktabs}
\usepackage{diagbox}
\usepackage{xcolor}
\newcommand{\de}{\mathrm{d}}
\newcommand{\ph}{\mathrm{Ph}}
\newcommand{\eps}{\epsilon}

\begin{document}
\title{Origins of Slow Magnetic Relaxation in Single-Molecule Magnets}

\author{Lei Gu}
%\author{Jie Li}
%\thanks{}
\affiliation{Department of Physics and Astronomy, University of California, Irvine, California 92697, USA}
\author{Ruqian Wu}
\email[]{wur@uci.edu}
%\thanks{}
\affiliation{Department of Physics and Astronomy, University of California, Irvine, California 92697, USA}

\begin{abstract}
Exponential and power law temperature dependences are widely used to fit experimental data of magnetic relaxation time in single molecular magnets. We derived a theory to show how these rules arise from the underling relaxation mechanisms and to clarify the conditions for their occurrence. The theory solves the puzzle of lower-than-expected Orbach barriers found in recent experiments, and elucidates it as a result of the Raman process in disguise. Our results highlight the importance of reducing the rate of direct tunneling between the ground state doublet so as to achieve long time coherence in magnetic molecules. To this end, large spin and small transverse magnetic anisotropy can reduce magnitude of the transition operator, and rigid ligands may weaken the spin-phonon coupling in that they raise the energy of vibrational modes and better screen the acoustic phonons.
\end{abstract}
\maketitle

Advances in quantum computing and quantum sensing technologies rely on the synthesis of innovative materials with predesigned properties and thorough fundamental understandings of their behavior. The electron spins in quantum dots were first suggested as qubits by DiVincenzo~\cite{DiVincenzo2000}, as their quantized spin states can be controlled and measured with electromagnetic stimuli. To extend the decoherence time for quantum operation, a robust qubit should be well isolated from its environment, yet effective communication is still needed for information exchange with others. To this end, single-molecule magnets (SSMs) are regaining exceptional 
research interest for developing platforms of quantum computation and information storage~\cite{Luis2018,Gaita2019}, as their spin is mostly protected by 
organic ligands and the exchange interaction across them can be easily controlled by varying the distance, substrate or charge state. 
Nevertheless, SMMs have numerous vibrational modes that may couple to spin excitation and hence how to extend the relaxation time of 
spin states is a central issue for the practical applications. It is perceived that molecules with large zero field splitting  
(e.g. large magnetic anisotropy energy), or equivalently with wide magnetic hysteresis 
~\cite{Zadrozny2013,Blagg2013,Chen2016,Goodwin2017,Guo2018, Randall2018} may have slow magnetic relaxation
~\cite{Ishikawa2003,Ardavan2007,Magnani2010,Harman2010,Freedman2010,Zadrozny2011,Lucaccini2014,Gomez2014,Moseley2018,Rajnak2019}. However, the general guiding rule for the search of molecular qubits has not been established.

Typical SMMs are complexes that involve a magnetic center and organic backbones. Together with solvent molecules, they may form molecular crystals. Due to strong coupling between the spin and organic backbones, the local vibrational modes play important roles in the quantum behaviors of SMMs. The Jahn-Teller effect~\cite{Garcia2006} may arise from coupling between local modes and excitation doublets~\cite{Garcia2005}. The interaction between the local modes and acoustic modes may essentially change the energy spectra~\cite{Pae2013} and cause cooperative spin cross over~\cite{Palii2015}. For the spin-lattice relaxation, the development of ab initio spin dynamics simulation~\cite{Moreno2017,Lunghi2017} allows quantitative investigations of spin-local mode coupling and recovers experimental relaxation rates. While the Orbach regime can be well accounted by ab initio calculations, the establishment of power laws requires other factors~\cite{Goodwin2017}. It is well known that coupling to acoustic modes can render power laws~\cite{Shrivastava1983}. In molecule crystals, however, energies of the acoustic modes are low due to weak inter-molecular interactions, so that they are likely incapable of exciting spin states. The condition for the power laws is a fundamental problem that has not been clarified.

Although existing theories of nuclear spin assisted tunneling~\cite{Giraud2001,Ishikawa2005,Gomez2014,Chen2017}, dipolar interaction~\cite{Harman2010,Ding2018}, and spin-lattice interaction~\cite{Shrivastava1983,Abragam2012,Jarmola2012,Lunghi2017,Escalera2018,Donati2020} can explain some phenomena in magnetic relaxation of SMMs, there are decades long puzzles in this realm. One of them is the presence of two Orbach barriers in some observations~\cite{Harman2010,Watanabe2011}. Another one that is more prevalent~\cite{Harman2010,Freedman2010,Vallejo2012,Coca2013,Zhu2013,Fataftah2014,Pedersen2015,Novikov2015,Rechkemmer2016} and still receives increasing attention~\cite{Rajnak2019,Wang2019,Kobayashi2019} is the under-barrier relaxation, where the observed barrier is significantly lower than that set by the magnetic anisotropy.

%It is acknowledged that spin-lattice coupling is important for the magnetic relaxation in various temperature ranges. The direct process is the first order scattering, where transition between two spin states is accompanied by energy exchange with a single phonon. For a spin system with $H_{spin}=-|D|S_z^2$, direct transitions need to climb over the overall barrier $|D|S^2$, known as the Orbach process. Its characteristic time has the exponential temperature dependence, $\tau\propto e^{|D|S^2/k_BT}$. Accordingly, finding magnetic units with high barriers is believed to be the key to practical applications, and research in this direction has received enormous attention. In many observations, however, the barriers for spin excitations are mysteriously low and do not correlate to energy difference among spin states. Moreover, the effective barriers may even drop as temperature decreasesthe.

Here, we propose a theory of spin-lattice relaxation in SMMs by combining the Redfield equation~\cite{Breuer2007} and non-equilibrium Green’s function (NEGF) method~\cite{Stefanucci2013,Xu2008}. The Redfield equation is a microscopic master equation describing evolution of an open quantum system. Given a microscopic Hamiltonian, NEGF derivations are deductive and automatically include various relaxation processes in a unified manner. Using models with large zero field splittings and local vibrational modes, we show that the low barriers have nothing to do with the Orbach process, but arise from direct tunneling between the (pseudo) ground state doublet. In addition, it shows that power laws can only arise from direct tunneling, and involvement of spin excited states compromises these laws. These results highlight the importance of reducing the tunneling rate for the design of practical SMMs devices.

Casting the correlation functions in the Redfield equation~\cite{Breuer2007} into NEGF, phonon induced relaxation is governed by
\begin{equation}
\frac{\de}{\de t} \rho_S(t)=\\
\sum_{\omega, q} i G^<_{q}(\omega)\left[A_{q}(\omega)\rho_SA_{q}^{\dagger}(\omega)-\frac{1}{2}\{A_{q}^{\dagger}(\omega)A_{q}(\omega), \rho_S\}\right],
\label{reduced}
\end{equation}
where $\rho_S(t)$ is the density matrix of the open system, $G^<_{q}(\omega)$ is the lesser Green’s function for phonons, the curly {\color{blue} bracket denotes} the anti-commutation, and $A_{q}(\omega)$ is the transition operator for spin eigenstates. For the transition from state $n$ to $m$, the operator elements are $A_q^{ij}=a_{q}\delta_{im}\delta_{jn}$, and the energy in Eq.~(\ref{reduced}) is defined as $\omega=\omega_m-\omega_n$. $A_{q}^{\dagger}(\omega)=A_{q}(-\omega)$ represents the reverse transition. Subscript $q$ means that the transition is caused by coupling with the $q$th phononic degree of freedom, a single phonon for the first order spin-phonon coupling and a pair of phonons for the second order coupling.

To the quadratic order, the spin Hamiltonian of a SMM takes the form as $H_{spin}=-DS_z^2-E(S_x^2- S_y^2)$. Most of SMMs designed for slow magnetic relaxation are easy axial ones, and a strong easy axial magnetic anisotropy ($D\gg E$) results in an ideal parabolic Orbach barrier. As $E$ is non-zero, direct tunneling between the ground state doublet is possible and an energy splitting renders the doublet a pseudo one (explained later). Assuming dominance of the direct tunneling, relaxation pathways via the excited states can be neglected. Magnetic relaxation of the ground state is described by
\begin{equation}
\begin{cases}
\frac{\de}{\de t} M = -2p_u M,\\
p_u=\sum_{q}i|a_{q}|^2G^<_{q}(\omega),
\end{cases}
\end{equation}
where $p_u$ denotes rate of the upward transition from the ground state to the state slightly lifted.

\begin{figure}
\centering
\includegraphics[width=0.8\textwidth]{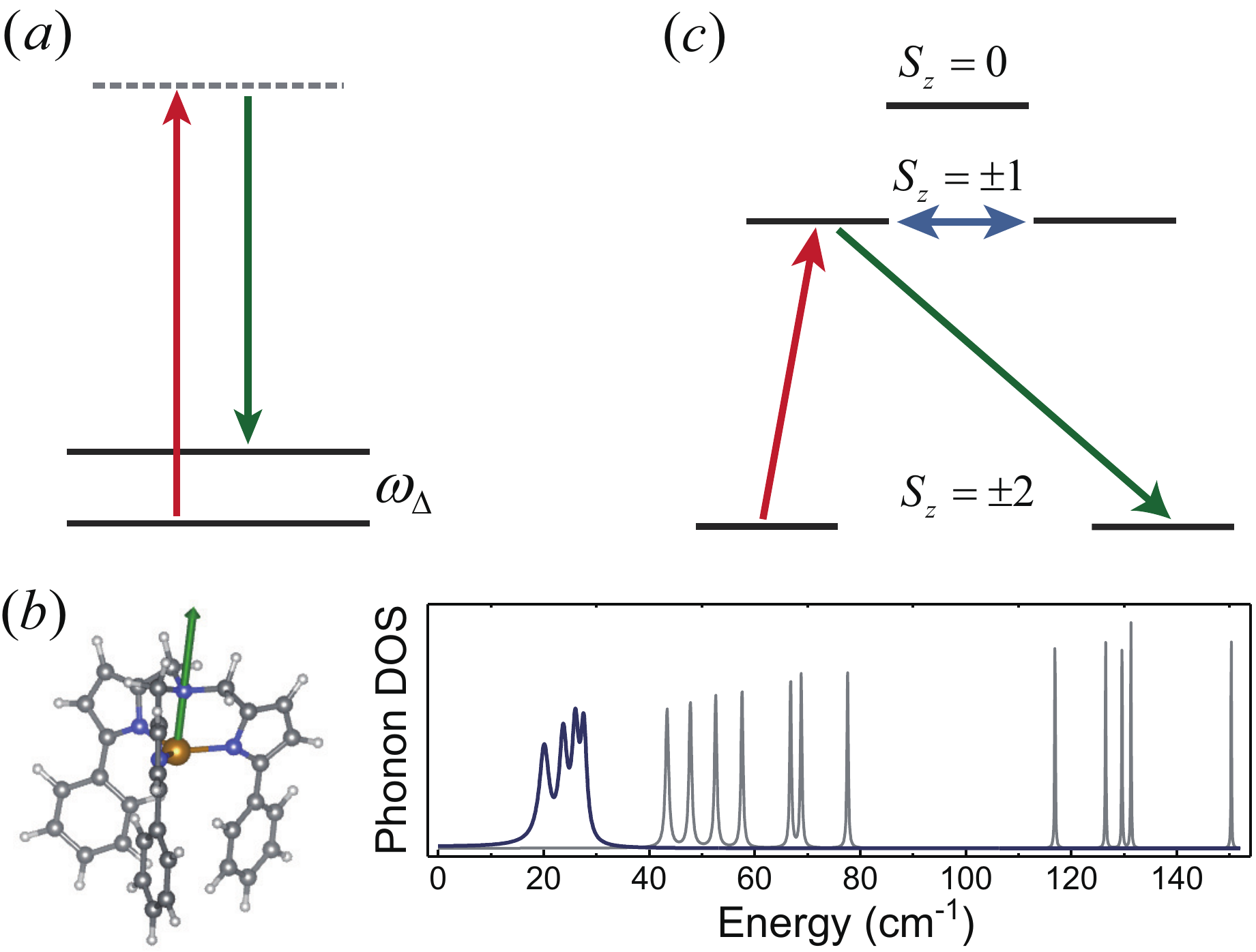}
\caption{(a) Coupling between the spin and the vibrational modes causes direct tunneling via the Raman process. (b) The [tpa$^{\ph}$Fe]$^{-1}$ is a $S=2$ molecule and possesses an easy axial magnetic anisotropy of $26$ cm$^{-1}$, and the four lowest vibrational mode energies are range from $20.1\sim27.6$ cm$^{-1}$. (c) Quadratic anisotropy can only yield $\Delta S_z = \pm 1, \pm2$ transitions. Due to divergent transition rate between degenerated states (here $S_z=\pm 1$), we can effectively take $|\pm 1\rangle$ as a single state and the magnetic relaxation involves the upward and downward transitions.}
\label{example}
\end{figure}

Due to the strong axial magnetic anisotropy ($D\gg E$), the energy splitting (denoted by $\omega_{\Delta}$ hereafter) between the ground state doublet is very small. Lack of energy match implies that the direct process through energy exchange with a vibrational mode is unviable, and the second order processes are needed. They arise from the coupling $H_2 = \sum_{qq'}\frac{\partial^2H_{spin}}{\partial V_{q}\partial V_{q'}}V_{q}V_{q'}$, where $V_{q}$ denotes the momentum space displacement. Here, the pair $(q,q')$ should be taken as a single phononic degree of freedom, and its Green's function can be calculated using $G_{qq'}^<(\omega)=\frac{i\hbar}{2\pi}\int  d \omega'  G_{q}^<(\omega)G_{q'}^<(\omega-\omega')$, where $G_{q}^<(\omega)$ is the single phonon lesser Green's function. Accordingly, the upward transition rate can be derived as
\begin{align}
p_u\propto N(\omega)\iint \frac{\de\omega_{q}\de\omega_{q'}}{\omega_{q}\omega_{q'}}\sigma(\omega_{q})\sigma(\omega_{q'}) \{&[N(\omega_{q})+N(\omega_{q'})+1]\delta(\omega-\omega_{q}-\omega_{q'})\nonumber\\
+ &[N(\omega_{q})-N(\omega_{q'})]\delta(\omega+\omega_{q}-\omega_{q'})\},
\label{pu2}
\end{align}
with $\omega=\omega_{\Delta}$ specifying the energy gain and $\sigma(\omega_{q})$ denoting the phonon DOS. Due to the inter-molecular interactions, the phonon DOS is not summation of delta functions, but has Lorentzian peaks around the mode energies $\omega_{\alpha}$ (e.g., see Fig.~\ref{example}(b)). By energy conservation, we can identify the first term as the double phonon process whereby two phonons are absorbed, and the second terms as the Raman process whereby a phonon is absorbed ($\omega_{q'}$) and a phonon of lower energy is emitted ($\omega_{q}$).

In the double phonon processes, energy summation of two phonons should match the transition energy, so they are also unviable due to the energy conservation. What matters are the Raman processes (Fig.\ref{example}(a)), which are represented by the second term in Eq.~(\ref{pu2}). Since $\omega=\omega_{\Delta}\ll 1$ cm$^{-1}$, $\omega_{q}-\omega_{q'}=\omega_{\Delta}$ implies that $\omega_{q}, \omega_{q'}$ are close. Namely, the absorbed and emitted phonons should be around the same Lorentzian DOS peak. Carrying out the integral with respect to the Lorentzian peak at $\omega_{\alpha}$, we obtain
\begin{equation}
\tau^{-1}=2p_u\simeq\frac{4\omega_{\alpha} \Gamma_{\alpha}|a_{\alpha}|^2}{(\omega_{\alpha}^2\omega_{\Delta})^2+(2\omega_{\alpha}\Gamma_{\alpha})^2}e^{-\omega_{\alpha}/k_BT},
\label{result}
\end{equation} 
where $\Gamma_{\alpha}$ is the broadening width, and $a_{\alpha}$ an overall alias of $a_{qq'}$ for $\omega_{q},\omega_{q'}\approx \omega_{\alpha}$.

Eq.~({\ref{result}}) indicates a vibronic barrier equal to the mode energy and explains the under-barrier relaxation. Due to the exponential form, it is likely that one or several of the lowest vibrational modes strongly coupling to the spin dominate the process. As a concrete test of the result, we calculated the vibration spectrum of [tpa$^{\ph}$Fe]$^{-1}$ (Fig.~\ref{example}(b)), a typical SMMs with slow magnetic relaxation~\cite{Harman2010}. Its four lowest vibrations fall in the range  $20.1\sim27.6$ cm$^{-1}$, followed by a much higher one at $43.4$ cm$^{-1}$. The transition rate $p_u$ in Eq.~({\ref{result}}) is summation over these vibrational modes, and leads to an effective barrier $20.1<E_{eff}<27.6$ cm$^{-1}$. This value is in accordance with the observed barrier of $26(2)$ cm$^{-1}$~\cite{Harman2010}, while the magnetic barrier of $3D=78$ cm$^{-1}$ is too high.

Based on Eq.~(\ref{result}), the puzzle of barrier lowering can also be explained. The ratio of transition rates for two DOS peaks reads  $p_u(\omega_{\alpha})/p_u(\omega_{\beta})\propto e^{(\omega_{\beta}-\omega_{\alpha})/k_BT}|a_{\alpha}|^2/|a_{\beta}|^2$. A lower vibrational mode with weaker spin-phonon coupling (say, $\omega_{\alpha}<\omega_{\beta}$ and $|a_{\alpha}|<|a_{\beta}|$) might have an advantage when temperature is low. As a result, a lower barrier characterizes the relaxation. For this lowering to be actually observed, however, a sizeable energy difference between the two modes is required. Otherwise, what shows up would be an averaged barrier. Moreover, the lower mode should have much weaker spin-vibration coupling, so that it is dominant only at low temperature rather than for all temperatures. These requirements explain why this barrier lowering is much less prevalent than observation of the under-barrier relaxation.

Applying Eq.~({\ref{pu2}}) to the Orbach process, we can see why the spin dynamics simulation in Ref.~\cite{Goodwin2017} cannot yield power laws. Without compromising the  physical essence, we take $S=2$ as an example. The Obarch process for $S=2$ spins with easy axial magnetic anisotropy follows the pathway in Fig.~\ref{example}(c). The magnetic relaxation rate also takes the form in Eq.~({\ref{pu2}}). The second order processes does not give rise to power laws, since the factor $N(3D)$ sets the dominant time scale $\tau=\tau_{0}e^{3D/k_BT}$, and the integration part only modifies the factor $\tau_{0}$. Carrying out the integral in Eq.~(\ref{pu2}) with respect to certain dominant DOS peaks, we have $\tau_{0}\propto T^0$ (temperature independent), an imperceptible modification. When the acoustic phonons are considered, one may have $\tau_{0}\propto T^{-1}$, which is still an insignificant modification compared to the exponential form itself. Clearly, we cannot obtain power laws for large zero field splitting, even if the second order processes and acoustic phonons are considered.

Going back to the direct tunneling by changing $3D$ to $\omega_{\Delta}$, and considering the coupling between spin and acoustic phonons, the standard derivations for the power laws are applicable, as the small transition energy $\omega_{\Delta}$ makes expansion w.r.t. $\omega_{\Delta}/k_{B}T$ and the Debye integral legitimate. While these standard results for small energy splittings are well known, the unviability to generate power laws for large zero field splittings appears to be not well aware of, and the community is puzzled on the origins of these relations~\cite{Goodwin2017, Lunghi2017}. This unviability indicates a correspondence between emergence of power laws and dominance of the direct tunneling. That is, upon observing the power laws, one can safely infer the dominance of the direct tunneling.

This correspondence has direct implication for the practical design of SMM devices. In the regime of exponential dependence, the relaxation time can be dramatically lengthened with small temperature reduction. The transition point from the exponential law to the power laws is the sweet point of long relaxation time at high temperature. For this reason, magnetic hysteresis usually cooccurs with the dominance of power law dynamics~\cite{Goodwin2017,Guo2018,Randall2018}, and this regime is the most suitable one for practical applications of SMMs. While the large Orbach barriers and wide molecular magnetic hysteresis in recent dysprosocenium SMMs~\cite{Goodwin2017,Guo2018,Randall2018} are appealing and receive lasting attentions, occurring in the power law regime, the broad magnetic hysteresis is due to the small direct tunneling rate, instead of the Orbach barrier. This calls for attention to reduce the tunneling rate besides the obsession on super large Orbach barriers.

\begin{figure}
\centering
\includegraphics[width=0.8\textwidth]{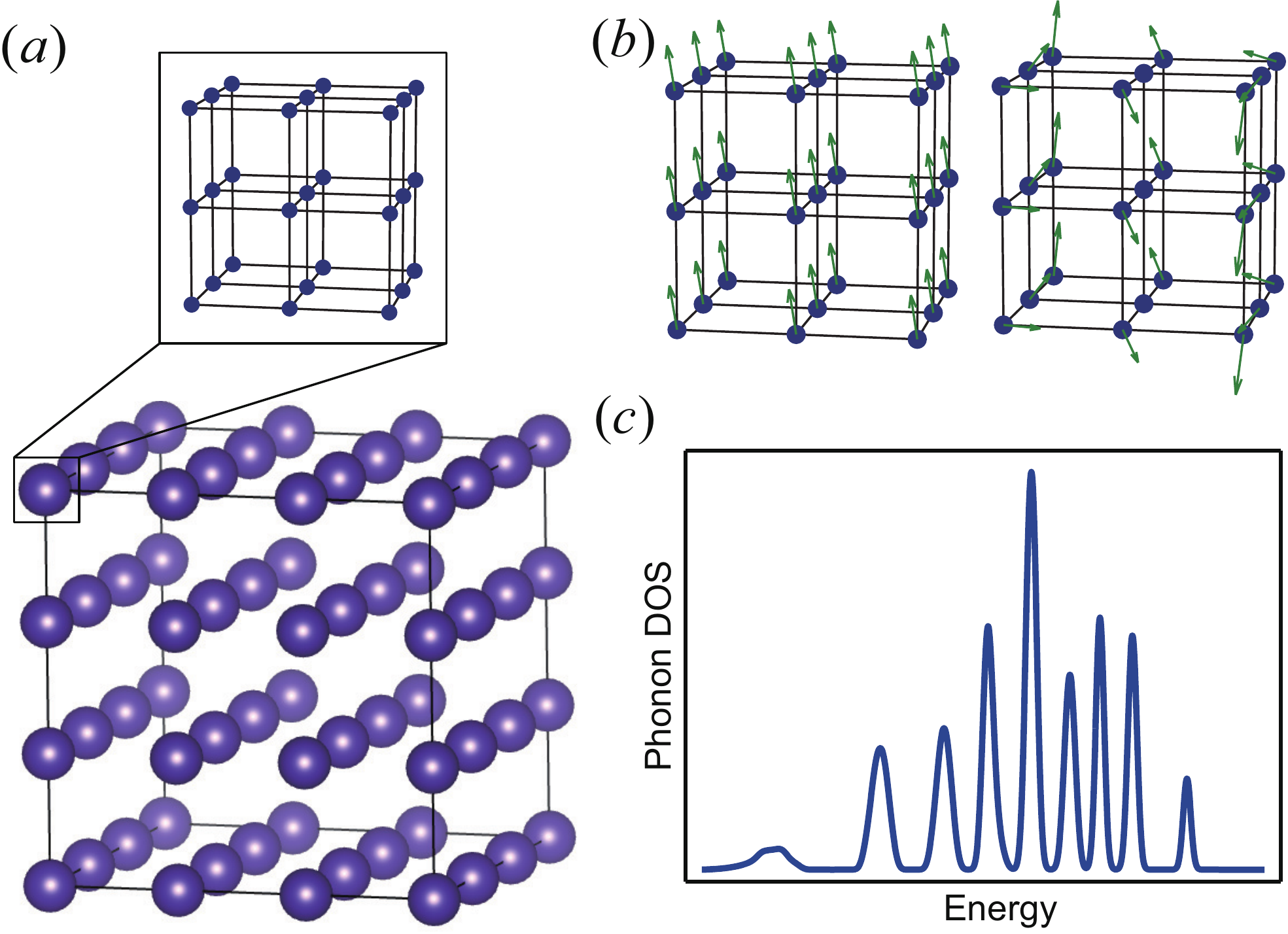}
\caption{(a) A 3D harmonic oscillator, where each cell is a $3\times3\times3$ cubic molecule. The inter-molecular force constant is one order smaller than that of intra-molecular interaction. (b) Relative atomic motion within a molecule is small for acoustic phonons (left) and significant for optical phonons (right). (c) The phonon spectrum consists of low energy Debye phonons and optical phonons resulting from broadening of local vibrational modes.}
\label{superlattice}
\end{figure}

Before systematic remarks on how to reduce the rate of direct tunneling, we need to address the question why the vibronic barrier can be raised, provided the optical phonons (vibrational modes) have much higher energies than the acoustic ones. In other words, why the acoustic phonons cannot always dominate, albeit they are energetically more accessible. To illustrate the reason, we calculated phonon modes of a 3D harmonic oscillator where the inter-molecular interaction is assumed to be one order smaller than the intra-molecular interaction, i.e., $k_{intra}=10 k_{inter}$. Fig.~\ref{superlattice}(b) shows atomic motions for an acoustic phonon (left) and optical phonon (right) with a momentum $(\frac{\pi}{2a},\frac{\pi}{2a}, 0)$. As spin-phonon coupling essentially represents the variation of electronic states due to atomic displacements, small relative motions in acoustic modes imply weak phonon-spin coupling. Moreover, it can also be seen that the DOS of the acoustic phonons is small too. Due to the weak coupling and small DOS compared to the optical phonons, the acoustic phonons can only be dominant at low temperature, when the high energy optical phonons are quite hard to access.

\begin{figure*}[t]
\centering
\includegraphics[width=0.9\textwidth]{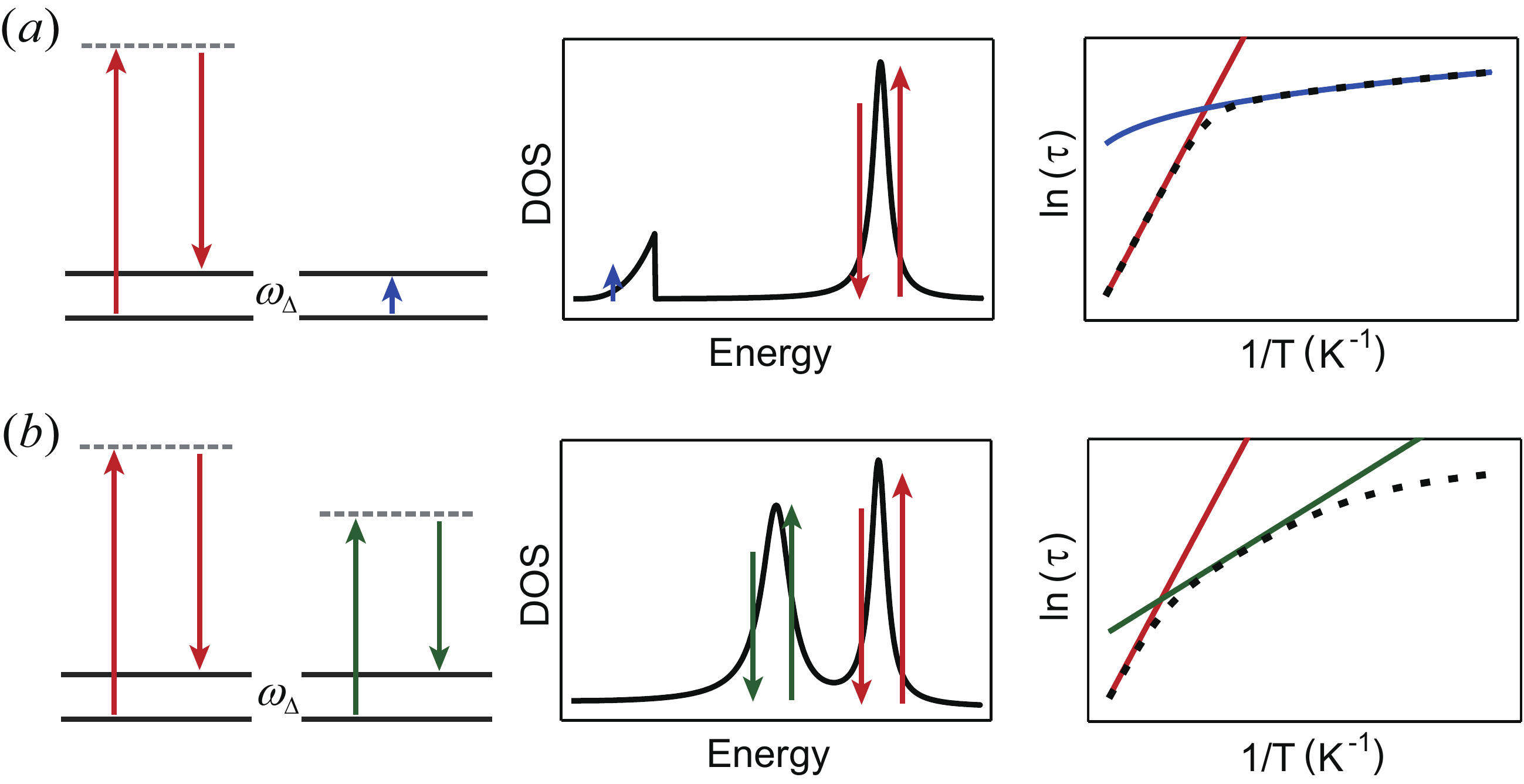}
\caption{Left, schematics of the relaxation processes; Middle, phonons responsible for the processes; Right, the corresponding relaxation-time-temperature dependences. (a) At high temperature, the Raman process due to phonons resulted from broadening of a vibrational mode dominates the relaxation, characterizing an exponential relation with energy of the mode as the relaxation barrier. (b) A vibrational mode with lower energy and weaker spin-phonon coupling could be dominant when temperature is reduced, and turns the relaxation barrier to a lower one.}
\label{dependences}
\end{figure*}

With all these understandings, we can relate our theory to the measured curves. With estimation on typical parameters, we produced the curves in Fig.~{\ref{dependences}}. The color code marks the correspondence among the relaxation processes, the phonons in charge, and the resultant relaxation time curves. Fig.~\ref{dependences}(a) gives the generic curve in most experimental observations, and Fig.~{\ref{dependences}}(b) is the case when two barrier are observed~\cite{Harman2010,Watanabe2011}. Here, in the low temperature range, we assume dominance of direct process ($\tau\propto T^{-1}$) for the acoustic phonons. Depending on the first and second order spin-phonon coupling strength, the Raman or double phonon processes may be dominant and raise other power laws.

To this point, the reasons for the slow magnetic relaxation are clear. At very high temperatures, the spin excited states are well accessible, and the Orbach process dominates. In this regime, the high Orbach barrier is responsible for the slowness, which is our conventional understanding. As discussed above, however, the Orbach regime may not be the best for practical application. The direct tunneling regime provides a better trade-off between long relaxation time and high temperature, and could be our major concern. Besides, because only the ground state doublet is involved in this regime, it makes a clear two state qubit. As for reducing the rate of direct tunneling, we may use SMMs with large spins and small transverse magnetic anisotropy $E$, and design more rigid backbones for the magnetic center.

The first principle can be understood with perturbation theory. For a spin with $E=0$, the ground states are two degenerated states consisting of $|\pm S\rangle$. Quadratic spin terms cannot yield any transition for $S>1$. The direct tunneling becomes possible, when $E(S_x^2-S_y^2)$ mixes state $|m\rangle$ ($-S<m<S$) into the ground state doublet, with mixing proportions $\propto E^{(m\mp S)/2}$ for $|\pm S\rangle$. As a result, increasing $S$ leads to exponential reduction of the tunneling rate. It is this small tunneling rate that gives rise to the broad magnetic hysteresis in those large spin ($S=15/2$) dysprosocenium SMMs~\cite{Goodwin2017,Guo2018,Randall2018}. As these molecules almost reach the limit of atomic angular momentum, reducing the $E$ value is a direction for further progress. Besides those magnetic engineerings, enclosing the magnetic center with more rigid backbones may better screen the spin from the acoustic phonons, and moreover, make the vibrational modes harder to access.

In conclusion, we have demonstrated how the direct tunneling between the ground state doublet gives rise to the observed temperature dependences of magnetic relaxation time. In particular, we found that the tunneling due to a vibrational mode can yield exponential temperature dependence, raising a relaxation barrier characterized by energy of the mode. Reasons for the slowness and hysteresis are systematically clarified and suggestions for improvements are provided. We proceeded with the problem of magnetic relaxation based on microscopic Hamiltonian and the density operator. This theory is fully quantum and may apply to describe general magnetic decoherence processes. The formulations are readily amenable for ab inito calculation for diverse quantum magnetic systems such as magnetic impurities, molecules and atoms.

The work is supported by the Department of Energy (grant No. DE-SC0019448) and computing time by NERSC.
%In the Orbach regime the relaxation time decreases exponentially with temperature, that is, small temperature decrease significantly slows down the relaxation. The transition point to the power law regime corresponds to the optimal trade-off between high temperature and slow magnetic relaxation. Even for SMMs bearing high barrier real Orbach relaxation, the power law regime is the most promising for practical applications. The clarification that power laws indicate dominance of direct tunneling calls for more attention on reducing the tunneling rate. In the regard that heavy lanthanide cations almost reach the limit of large molecular spin, diminishing transverse and high order magnetic anisotropies and lifting energies of the vibrational modes are of practical importance for further progress.

%\section*{acknowledgements}

%\bibliography{references}

\pagebreak
\widetext
\begin{center}
\textbf{\large Supplemental Materials}
\end{center}
\section{Fundamentals}
\subsection{The NEGF form of Redfield euqation for spin-lattice relaxation}
The Redfield equation is given by~\cite{Breuer2007}
\begin{equation}
\begin{cases}
\frac{d }{d t}\rho_S(t) =-i[H_{S},\rho_S(t)] + L(\rho_S(t))\\
H_{S} = \sum_{\omega,qq'}S_{qq'}(\omega)A^{\dagger}_{q}(\omega)A_{q'}(\omega)\\
L(\rho_S) = \sum_{\omega,qq'}\gamma_{qq'}(A_{q'}(\omega)\rho_S(\omega)A^{\dagger}_{q}(\omega)-\frac{1}{2}\{A^{\dagger}_{q}(\omega)A_{q'}(\omega),\rho_S\}),
\end{cases}
\end{equation}
where the coupling factors are defined as
\begin{equation}
\begin{cases}
S_{qq'}(\omega) = \frac{1}{2i}\left[\Gamma_{qq'}(\omega)-\Gamma^{*}_{q'q}(\omega)\right]\\
\gamma_{qq'} = \Gamma_{qq'}(\omega)+\Gamma^{*}_{q'q}(\omega)
\end{cases}
\end{equation}
with
\begin{equation}
\Gamma_{qq'} = \int_{0}^{+\infty}ds\frac{e^{i\omega s}}{\hbar^2}\langle B^{\dagger}_{q}(t)B_{q'}(t-s)\rangle.
\end{equation}
Since we assume a phonon bath in equilibrium, $H_S$ is time-independent and can be absorbed into the unperturbed Hamiltonian (with a small level normalization of the spin system), and the relaxation is attributed to $L(\rho_S)$, which is our concern.

On the above $A, B$ are operators for the systems of interest and environmental degrees of freedom, respectively. That is, the interaction Hamiltonian takes the form 
\begin{equation}
H_{int} = \sum_{q}A_{q}B_{q},
\end{equation}
which can be cast into
\begin{equation}
H_{int} = \sum_{q}A_{q}(\omega)B_{q},
\label{project}
\end{equation}
by defining the transition operators
\begin{equation}
A_{q}(\omega) = \sum_{\eps'-\eps=\omega}\Pi(\eps)A_{q}\Pi(\eps')
\label{trans}
\end{equation}
where $\eps, \eps'$ denote energies of the target and source state, respectively. $\Pi(\eps)$ is the projection operator onto the state of energy $\eps$, and due to their operations only one element of $A_q(\omega)$ is nonzero.

The first order spin-phonon coupling is given by
\begin{equation}
H_{spin-vib} = \sum_{q}\frac{\partial H_{spin}}{\partial V_{q}}V_{q},
\end{equation}
where $ H_{spin}$ is the spin Hamiltonian and $V_{q}$ is displacement defined as
\begin{equation}
V_{q} =  \sqrt{\frac{\hbar}{2\omega_{q}}}(b_{q} + b^{\dagger}_{-q})
\end{equation}
with $b_{q}, b^{\dagger}_{q}$ the annihilation and creation operators for phonon $q$. For a phonon bath in equilibrium, $\Gamma_{qq'}=\delta_{qq'}\Gamma_{qq}$, for which we use one index for notation in the following.

According to the definition of the lesser Green's function~\cite{Xu2008,Stefanucci2013}, 
\begin{equation}
G_q^<(t,t') = -\frac{i}{\hbar}\langle V_q^{\dagger}(t')V_q(t)\rangle
\end{equation}
$\Gamma_{qq'}$ is equivalent to
\begin{equation}
\Gamma_{q} =\frac{i}{\hbar} \int \theta(t-t') \de(t'-t)e^{i\omega(t-t')}G_q^<(t',t),
\end{equation}
and similarly
\begin{equation}
\Gamma_{q}^* = \frac{i}{\hbar}\int \theta(t'-t) \de(t'-t)e^{i\omega(t-t')}G_q^<(t',t).
\end{equation}
In steady states, only the time difference matters, i.e., $G_q^<(t',t)=G_q^<(t'-t)$. Identity $\theta(t-t')+\theta(t'-t)=1$ implies that $\gamma_{q}(\omega)$ is nothing but the Fourier transformation,
\begin{equation}
\gamma_{q}(\omega) = \frac{i}{\hbar} G_{q}^<(-\omega).
\end{equation}
The minus sign in $G_{q}^<(-\omega)$ arises due to the reverse order of $t,t'$ in $G_q^<(t'-t)$ and in the phase factor $e^{i\omega(t-t')}$.

%Then $S_{q}$ can be written as
%\begin{equation}
%S_{q} = \frac{1}{2i\hbar^2}\left(\int dt e^{i\omega t}\theta(t-t') V^{\dagger}_{q}(t)V_{q}(t')-\int dt e^{i\omega t}\theta(t-t') V_{q}(t')V_{q}(t)\right),
%\end{equation}
%which is exactly definition of the retarded Green's function differed by a factor. Therefore,
%\begin{equation}
%S_{q}(\omega)=\frac{1}{2\hbar}G^r_q(\omega)
%\end{equation}
According to Eq.~(\ref{trans}), for transition between two spin states $n\rightarrow m$, elements of $A_{q}(\omega)$ take the form $A_{q}^{ij}(\omega)=a_{q}\delta_{in}\delta_{jm}$ with $\omega=\omega_{nm}=\omega_n-\omega_m$ and $a_q=\langle m|A_q|n\rangle$. We see here that $\omega$ is defined as $\omega_{source}-\omega_{target}$. By redefining it as $\omega_{mn}=\omega_m-\omega_n=-\omega_{nm}$, the minus sign in $G_{q}^<(-\omega)$ can be cancelled and we achieve the main text Eq.~(1). Considering a pair of phonons as a single degree of freedom, these arguments also apply to the second order spin-phonon coupling.

\subsection{Spin-phonon coupling Hamiltonian}
To the second order of magnetic anisotropies, a spin Hamiltonian takes the general form
\begin{equation}
H_{spin}=\vec{S}\mathbf{D}\vec{S},
\end{equation}
where $\mathbf{D}$ is an $3\times3$ matrix and $\vec{S}=(S_x, S_y, S_z)$. The phonon-spin coupling is variation of $\mathbf{D}$ caused by atomic displacements (taking the first order coupling as instance),
\begin{equation}
H_{sp-ph} = \sum_{i_s}\frac{\partial H_{spin}}{\partial V_{i_s}}\Bigg|_{V_{i_s}=0}V_{i_s}.
\end{equation}
Here, $i$ indexes atoms and $s = x,y,z$. Since the dynamical matrix for phonons are mass normalized, it is convenient to use mass normalized displacement, i.e., here $V_{i}=\sqrt{m_i}U_{i_s}$ with $U_{i_s}$ denoting atomic displacements. The Fourier transform of $V_{i_s}$ is given by
\begin{equation}
V_{i_s} = \sum_{q_b}\frac{e^{iq\cdot R_{i}}}{{\sqrt{N}}}\epsilon_{i_s}^{q_b} V_{q_b},
\end{equation}
where $V_{q_b}$ is magnitude of phonon vibration with $q$ denoting momentum and $b$ the branches. $R_{i}$ is the equilibrium position of the $i$th atom and $\epsilon_{i_s}^{q_b}$ elements of the polarization tensor (eigenvectors of the dynamical matrix). Due to momentum conservation, standing waves are excited or absorbed. This implies $V_{-q_b}=\pm V_{q_b}$. Together with $\epsilon_{i_s}^{q_b}=(\epsilon_{i_s}^{-q_b})^*$, summation over momentum pairs $(q,-q)$ leads to
\begin{equation}
H_{sp-ph} = \sum_{q_b}\left[\sum_{i_s}\frac{2\mathrm{Re}[e^{q\cdot R_i}\epsilon_{i_s}^{q_b}]}{\sqrt{N}}\frac{\partial H_{spin}}{\partial V_{i_s}}\Bigg|_{V_{i_s}=0}\right] V_{q_b}^{+}+i\sum_{q_b}\left[\sum_{i_s}\frac{2\mathrm{Im}[e^{q\cdot R_i}\epsilon_{i_s}^{q_b}]}{\sqrt{N}}\frac{\partial H_{spin}}{\partial V_{i_s}}\Bigg|_{V_{i_s}=0}\right] V_{q_b}^{-},
\label{coupling}
\end{equation}
where $V^{\pm}$ are displacements for the two branches of standing waves. A complex is a rather independent entity and only moments of the atoms around the magnetic center atom can effectively affect the spin.  We should take the magnetic center as reference, and what really takes effect is the relative value $\epsilon_{i_s}^{q_b}-\epsilon_{o_s}^{q_b}$, where $o$ indexes the magnetic center. The expression in the bracket is what we denote by $\partial H_{spin}/\partial_{V_q}$, from which the spin-phonon coupling is roughly proportional to intra-molecule motions.

\subsection{The 3D harmonic oscillator}
The force constants are set as $k_{intra}=10k_{inter}$, and the molecular distance is that of edge length of the molecule. Only the nearest neighbour interaction are considered. What we mean by atomic motion is the quantity $\mathrm{Re}[e^{q\cdot R_i}\epsilon_{i_s}^{q_b}]$, for which the magnetic center is chosen as the origin, $R_o=(0,0,0)$.

\subsection{Ab initio calculations} The vibrational modes and the spin Hamiltonian of [tpa$^{\ph}$Fe]$^{-1}$ were calculated with the ORCA package~\cite{Neese2018}. The adopted basis sets are a def2-TZVP basis set for Fe and N, def2-SVP for C and H and a def2-TZVP/C auxiliary basis set for all the elements, which have been shown to work well for this molecule~\cite{Lunghi2017}. Calculation of the vibrational modes is implemented at the DFT level with the PBE functional~\cite{PBE}. The lowest modes have energies of $20.1, 23.7, 26.0, 27.6$ cm$^{-1}$, and the next mode is at $43.4$ cm$^{-1}$. CASSCF~\cite{Atanasov2015} calculation with a (6,5) active space was carried out to obtain the spin Hamiltonian, which gives $D=30.3$ cm$^{-1}$ and $E/D=0.005$. We adopted the experimental value $D=26$ cm$^{-1}$ and the calculated ratio.

\subsection{Typical energy scales}
The Debye frequencies for ordinary materials are in the order $\lesssim 100$ cm$^{-1}$. Assuming the inter-molecular interaction is one order weaker than the chemical bounds, and mass of a complex is one order larger than an ordinary atom, from the dynamical matrix we estimate the Debye frequency for the inter-molecule motions as $\omega_D\lesssim 10$ cm$^{-1}$. According to experimental measurement~\cite{Ding2018} and ab initio calculation, the splitting between the ground state doublet is in the range $10^{-5}\sim 10^{-3}$ cm$^{-1}$. Thus $\omega_{\Delta}\ll \omega_{D}$. Under a magnetic field of $1500$ Oe, it is in the order $\sim 0.1$ cm$^{-1}$, still $\omega_{\Delta}\ll \omega_{D}$.

When any interaction with a vibration mode is considered, the Dyson equation for the mode gives (throughout this work Greek letters are used to index unperturbed vibrational modes)
\begin{equation}
G^r_{\alpha}(\omega) = \frac{1}{\omega^2-\omega_{\alpha}^2-\Sigma^r_{\alpha}(\omega)},
\end{equation}
where $\Sigma^r_{\alpha}(\omega)$ is the retarded self-energy. Absorbing the real part of $\Sigma^r_{\alpha}(\omega)$ into $\omega_{\alpha}$ and according to
\begin{equation}
\sigma_{\alpha}(\omega) = -\frac{2\omega_{\alpha}}{\pi}\mathrm{Im}[G^r_{\alpha}(\omega)],
\end{equation}
the DOS $\sigma(\omega)$ takes the Lorentzian form
\begin{equation}
\sigma(\omega)=\sum_{\alpha}\sigma_{\alpha}(\omega)=\sum_{\alpha}\frac{1}{\pi} \frac{4\omega_{\alpha}\Gamma_{\alpha}}{(\omega^2-\omega_{\alpha}^2)^2+\Gamma_{\alpha}^2},
\label{dos}
\end{equation}
with $\Gamma_{\alpha}=\mathrm{Im}[\Sigma^r_{\alpha}(\omega)]$. From the perturbational perspective, high order interactions with the Debye phonons are needed to calculate the broadening, since the Debye frequency is small and multiple Debye phonons are required to reach the energy level of vibrational modes. There is no simple estimation for the value. In Ref.~\cite{Goodwin2017}, broadening  $\sim 10$ cm$^{-1}$ nicely accounts for the experimental data. As higher order interactions are need for broadening of vibrational modes with higher energies, the broadening width tends to decrease with energy. In the main text plots, such narrowing is introduced to reflect this trend.

\subsection{Elements of the transition operator}
As presented in the first section, the coupling Hamiltonian in the Redfield equation takes the form
\begin{equation}
H_{int} = \sum_{q}A_{q}(\omega)B_{q},
\end{equation}
where $\omega=E_{target}-E_{source}$. It is the most convenient to do the calculations in the basis of eigenstates. For transition $n \rightarrow m$, according to the definition
\begin{equation}
A_{q}(\omega) = \sum_{\eps-\eps'=\omega}\Pi(\eps)A_{q}\Pi(\eps'),
\label{casted}
\end{equation}
where the project operators take the form $\Pi(\eps')=\delta_{nn}, \Pi(\eps)=\delta_{mm}$, we have the elements
\begin{equation}
A_q^{ij}(\omega_{mn}) = a_q\delta_{im}\delta_{jn}
\end{equation}
with 
\begin{equation}
a_q = \langle n| A_q| m \rangle.
\end{equation}
Here, $A_q$ is the original coupling operator when we write the coupling Hamiltonian in the form
\begin{equation}
H_{int} = \sum_{q}A_{q}B_{q}.
\end{equation}
The two forms of Hamiltonian are equivalent, since the summation over $\omega$ in Eq.~(\ref{casted}) implies the complete relation.

\section{Orbach process}
\subsection{The master equation}
For the effective two step relaxation illustrated in the main text Fig.~1(c), the evolution of the density matrix elements is given by
\begin{align}
d\rho_{22} &= \rho_{11}p_2^1-\rho_{22}p_1^2\\
d\rho_{-2-2} &= \rho_{11}p_{-2}^1-\rho_{-2-2}p_1^{-2}.
\end{align}
Here, we effectively take the $|1\rangle, |-1\rangle$ as a single states. The $p^1_{-2}, p^{-2}_{1}$ are actually $p^{_1}_{-2}, p^{-2}_{-1}$. With negligible transverse magnetic anisotropy $E$, the system is approximately symmetric about $|m\rangle$ and $|-m\rangle$, so we have $p_{-2}^1\equiv p_{-2}^{-1}\simeq p_{2}^{1}$, $p_{1}^{-2}\equiv p_{-1}^{-2}\simeq p_{1}^{2}$. Then, the above equations lead to evolution of the magnetization,
\begin{equation}
\begin{cases}
\frac{\de }{\de t} M =- p_u M,\\
M = S(\rho_{22}-\rho_{-2-2}),
\end{cases}
\end{equation}
where $p_u$ is an alias of $p_2^1$. The solution of these equations is
\begin{equation}
M = e^{-t/\tau}=e^{-p_u t}.
\end{equation}

\subsection{The direct process}
The direct process is attributed to the first order spin-phonon coupling $H_1=\sum_{q}\frac{\partial H_{spin}}{\partial V_{q}}V_{q}$ (i.e., $A_q=\frac{\partial H_{spin}}{\partial V_{q}}$) with $V_{q}$ denoting the displacement. As shown in the main text, $p_u$ is given by
\begin{equation}
p_u = \sum_{q}iG^<_{q}(\omega) |a_q|^2/\hbar,
\label{rate1}
\end{equation} 
where $\omega=3D$. Throughout this paper, Greek letters index the unperturbed vibrational modes, and $q,q',q'''$ are used for phonon indices. When the finite lifetime is {\bf not} considered, the lesser Green's function of phonons is given by~\cite{Xu2008}
\begin{equation}
G^<_{q}(\omega)=
-\sum_{q}\frac{i \pi}{\omega_q} \{\delta(\omega-\omega_q)N(\omega_q)+\delta(\omega+\omega_q)[N(\omega_q)+1]\},
\label{single}
\end{equation}
where $N(\omega)$ is the Bose-Einstein distribution. Writing the summation over $q$ as integration with respect to phonon DOS, relaxation time due to the direct process can be obtained as
\begin{equation}
\tau=\frac{1}{p_u}=\frac{\hbar\omega_{q}}{N(\omega_q)\pi \sigma(\omega_q)|a_q|^2}\simeq e^{3D/k_BT}\frac{\hbar3D}{\pi \sigma(3D)|a_q|^2}.
\label{direct}
\end{equation}
In the last step, $\omega_q=3D$ is substituted into, and the Bose-Einstein distribution is approximated by $N(3D)\simeq e^{-3D/k_BT}$ for $3D\gg k_BT$.  Here, in order to reach the concise expression, we assume an identical $|a_q|$ for all the phonons with $\omega_q=3D$. In ab initio calculation of the relaxation time, variation of $a_q$ can be accounted by numerically implementing the summation.

\subsection{Effect of finite phonon lifetime}
It is known that when lifetime of phonons is considered, the relaxation barrier is reduced a little~\cite{Lyo1972}. For SMMs, it was argued that resonance could lower the barrier to observed values~\cite{Lunghi2017}. Based on NEGF results, here we reveal how a barrier reduction could possibly arise, and why such mechanism is suppressed in general. Since the decaying of phonons also involves two phonons, we use Raman and double phonon decaying for the terminologies. One should bear in mind that in this section the spin-phonon coupling is in the first order. The Raman and double phonon decaying refer to the interactions with other two phonons that broaden the phonon of interest  

When anharmonic interaction is considered, the retarded Green's function can be formally written as
\begin{equation}
G^r_{q}(\omega) = \frac{1}{\omega^2-(\omega_{q}+\Delta_{q}-i/\tau_q)^2},
\end{equation}
where $\Delta_{q}$ is an energy shift, and broadening $1/\tau_{q}$ is the inverse lifetime. Compared to the Dyson equation
\begin{equation}
G^r_{q}(\omega) =  \frac{1}{\omega^2-\omega_{q}^2-\Sigma^r_{q}(\omega)},
\end{equation}
in the perturbation domain $|\Delta_{q}-i\Gamma_{q}| \ll \omega_{q}$, we have relation
\begin{equation}
\frac{1}{\tau_{q}} \simeq -\frac{\mathrm{Im}[\Sigma^r_{q}(\omega_{q})]}{2\omega_{q}}.
\end{equation}
The real part of the self-energy can be absorbed into $\omega_{\alpha}$, and this imaginary part is our concern.

According to identity $G_q^<(\omega)=G_q^r(\omega)\Sigma_q^<(\omega)G_q^a(\omega)$, The lesser self-energy is needed to calculate $G_q^<(\omega)$. It also gives the imaginary part of the retarded self-energy by $2N(\omega)\mathrm{Im}[\Sigma_q^r(\omega)]=\mathrm{Im}[\Sigma_q^<(\omega)]$, and accordingly $G^r_{q}(\omega)$ and $G^a_{q}(\omega)=(G^r_{q}(\omega))^{\dagger}$ can be derived. Finally, the lesser Green's function is given by
\begin{equation}
G^<_{q}(\omega)=N(\omega)\frac{-2i\Lambda_{q}}{(\omega^2-\omega_{q}^2)^2+\Lambda_{q}^2},
\label{lessg}
\end{equation}
where $\Lambda_{q}=-\mathrm{Im}[\Sigma_{q}^r(\omega)]=-\mathrm{Im}[\Sigma^<_{q}(\omega)]/2N(\omega)$. The minus sign is purposefully added here for later notational simplicity.

\begin{figure}
\centering
\includegraphics[width=0.6\textwidth]{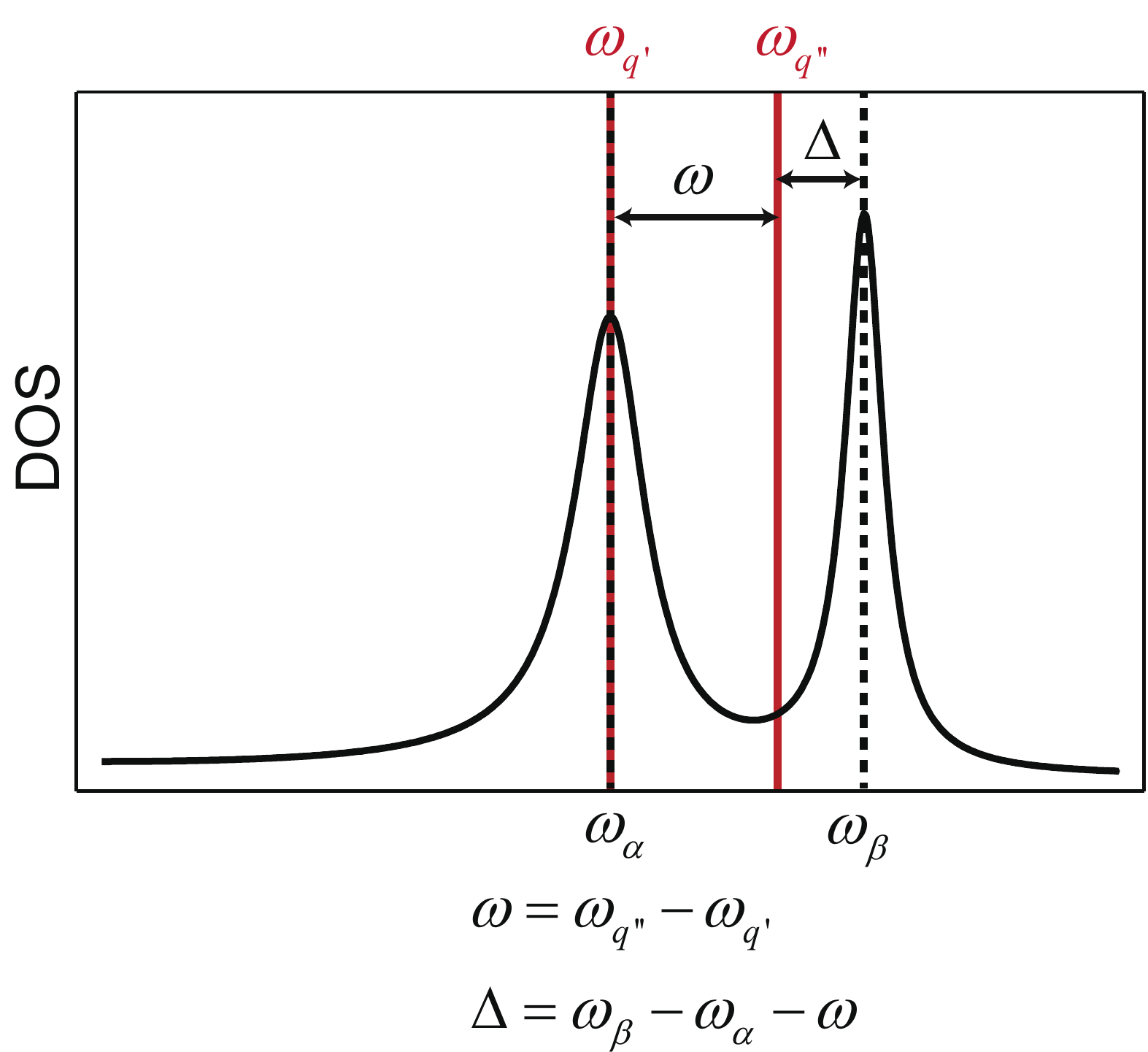}
\caption{Quantities in the integral for the Raman process. The transition energy $\omega$ set the energy difference between the absorbed and emitted phonons. The integral over $q',q''$ is significant when the deviation $\Delta$ from the resonance position is not bigger than broadening of the vibrational modes, that is, $\Delta\lesssim \Gamma_{\alpha}, \Gamma_{\beta}$.}
\label{integral}
\end{figure}

From the anharmonic interaction $H_{ah}=\sum_{qq'q''}\frac{\partial^3 P(\mathbf{V})}{\partial V_{q}\partial V_{q'}\partial V_{q''}}V_{q} V_{q'}V_{q''}$, the lesser self-energy can be derived as~\cite{Xu2008}
\begin{align}
\Sigma_{q}^<(\omega)=N(\omega)
\sum_{q'q''}\frac{-i\pi\hbar}{\omega_{q'}\omega_{q''}}|\phi(qq'q'')|^2\{&[N(\omega_{q'})+N(\omega_{q''})+1][\delta(\omega-\omega_q'-\omega_{q''}) -\delta(\omega+\omega_q'+\omega_{q''}))]\nonumber\\
+ &[N(\omega_{q'})-N(\omega_{q''})][ \delta(\omega+\omega_{q'}-\omega_{q''}) -\delta(\omega-\omega_{q'}+\omega_{q''})]\},
\label{sigma2}
\end{align}
where $\phi(qq'q'')=\frac{\partial^3 P(\mathbf{V})}{\partial V_{q}\partial V_{q'}\partial V_{q''}}$. Here, the frequency $\omega$ in the factor $N(\omega)$ can be negative. Noting $N(-|\omega|)+1=-N(|\omega|)$, Eq.~(\ref{sigma2}) can be cast into a simpler form
\begin{align}
\Sigma_{q}^<(\omega)=S(\omega)\sum_{q'q''}\frac{-i\pi\hbar}{\omega_{q'}\omega_{q''}}|\phi(qq'q'')|^2\{&[N(\omega_{q'})+N(\omega_{q''})+1]\delta(|\omega|-\omega_{q'}-\omega_{q''})\nonumber\\
+ &[N(\omega_{q'})-N(\omega_{q''})]\delta(|\omega|+\omega_{q'}-\omega_{q''})\},
\label{sigma2p}
\end{align}
where the second term is restricted to $\omega_{q''}>\omega_{q'}>0$ and $S(\omega)$ is defined as
\begin{equation}
S(\omega)=
\begin{cases}
N(|\omega|)+1 \ \ \omega<0,\\
N(\omega) \ \ \ \ \ \omega>0.
\end{cases}
\end{equation}
In the [tpa$^{\ph}$Fe]$^{-1}$ example, only the upward transition matters, i.e., $\omega> 0$. The summation over $q', q''$ is pronounced, when $\omega_{q'}, \omega_{q''}$ lie around two DOS peaks that satisfies $\omega_{\alpha}\pm\omega_{\beta}-\omega=\Delta$, with $\Delta\lesssim \Gamma_{\alpha}, \Gamma_{\beta}$. Signs $\pm$ correspond to the double phonon and Raman decaying processes, respectively. Taking the Raman process for instance, we can set $\omega_{q'}=\omega_{\alpha}$, and then it is clear that $\Delta$ represents deviation from the resonance position (Fig.~\ref{integral}).

Substituting the Lorentzian DOS functions $\sigma_{\alpha}(\omega_{q'}), \sigma_{\beta}(\omega_{q''})$ (cf. Eq.~(\ref{dos})) into Eq.~(\ref{sigma2p}), we get the integration form. To reach a compact expression, we assume the coupling strengths $|\phi(qq'q'')|$ are similar for the phonons around the two DOS peaks ($\omega_{q'}\approx \omega_{\alpha}, \omega_{q''}\approx \omega_{\beta}$), and denote the value by $|\Phi_{q\alpha\beta}|$. Since the contribution mainly comes from the peak areas, we can approximate $N(\omega_{q'})$ and $N(\omega_{q''})$ with $N(\omega_{\alpha})$ and  $N(\omega_{\beta})$, respectively. Then, with variable change $x=\omega-\omega_{q'}$, the integral reduces to (does not include constant factors)
\begin{equation}
I_{\pm} \simeq \int dx \frac{2\Gamma_{\alpha}}{(2\omega_{\alpha}x+x^2)^2+\Gamma_{\alpha}^2}\frac{2\Gamma_{\beta}}{[2\omega_{\beta}(\Delta\mp x)+(\Delta\mp x)^2]^2+\Gamma_{\beta}^2}.
\label{integrand}
\end{equation}
According to the residue theorem, the integral can be obtained as
\begin{align}
I_{\pm}=& i\frac{2\pi}{4}\left(\frac{1}{\pm\omega_{\alpha}\omega_{\beta}\Delta+i\omega_{\alpha}\Gamma_{\beta}+i\omega_{\beta}\Gamma_{\alpha}}-c.c. \right)\\
=&\frac{\pi(\omega_{\alpha}\Gamma_{\beta}+\omega_{\beta}\Gamma_{\alpha})}{(\omega_{\alpha}\omega_{\beta}\Delta)^2+(\omega_{\alpha}\Gamma_{\beta}+\omega_{\beta}\Gamma_{\alpha})^2}.
\end{align}
The signs $\pm$ do not matter here, that is, the Raman process and double process have the same expression (but correspond to different mode pairs $\omega_{\alpha}, \omega_{\beta}$ ). Adding the factors, we have
\begin{align}
\Sigma_{q}^<(\omega)=\frac{-4i\hbar N(\omega)(\omega_{\alpha}\Gamma_{\beta}+\omega_{\beta}\Gamma_{\alpha})}{(\omega_{\alpha}\omega_{\beta}\Delta)^2+(\omega_{\alpha}\Gamma_{\beta}+\omega_{\beta}\Gamma_{\alpha})^2}|\Phi_{q\alpha\beta}|^2\{&[N(\omega_{\alpha})+N(\omega_{\beta})+1]\nonumber\\
+ &[N(\omega_{\beta})-N(\omega_{\alpha})]\},
\end{align}
and $\Lambda_{q}=-\mathrm{Im}[\Sigma^<_{q}(\omega)]/2N(\omega)$ gives
\begin{align}
\Lambda_{q} \propto \frac{\omega_{\alpha}\Gamma_{\beta}+\omega_{\beta}\Gamma_{\alpha}}{(\omega_{\alpha}\omega_{\beta}\Delta)^2+(\omega_{\alpha}\Gamma_{\beta}+\omega_{\beta}\Gamma_{\alpha})^2}\{&[N(\omega_{\alpha})+N(\omega_{\beta})+1]\nonumber\\
+ &[N(\omega_{\alpha})-N(\omega_{\beta})]\}.
\label{broaden}
\end{align} 

%Here, we make the approximation $|\Phi_{qq'q''}|\approx|\Phi_{q\alpha\beta}|$ for $\omega_{q'}, \omega_{q''}$ around $\omega_{\alpha}, \omega_{\beta}$.

\begin{figure}
\centering
\includegraphics[width=0.6\textwidth]{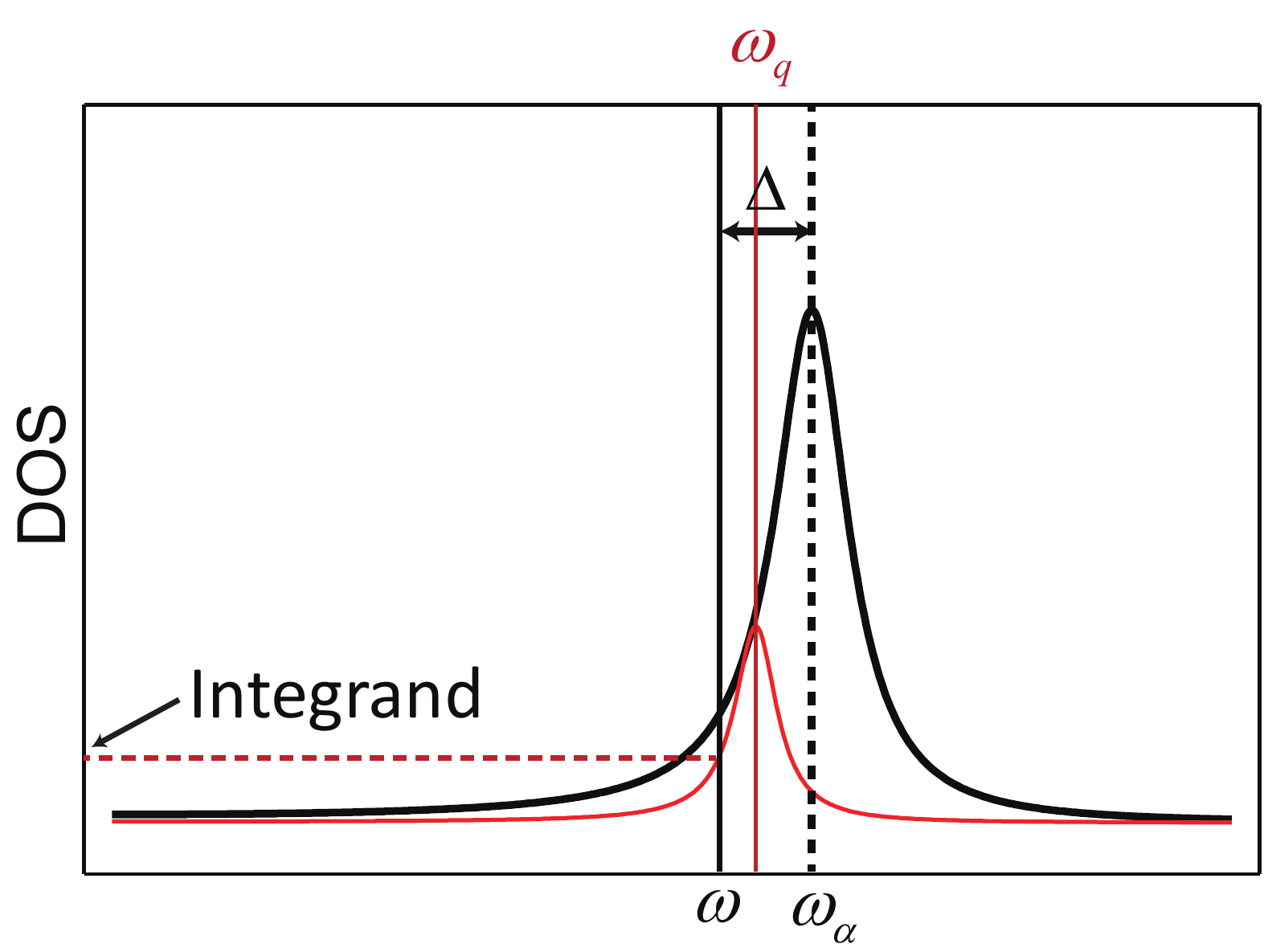}
\caption{Integrand of calculating the transition rate of direct process when finite lifetime of phonons is considered. As a relaxation process due the first order spin-phonon coupling, it is required in the first place that a vibrational mode has energy close to the transition energy $\omega=3D$. The DOS profile (bold black) due to broadening of vibrational modes serves as an envelope of the broadening function (red) due to finite lifetime. Denoting the deviation as $\Delta=\omega_{\alpha}-\omega$, we can have an expression of the integrand similar with Eq.~(\ref{integrand}).}
\label{lifetime}
\end{figure}

Substituting $G^<_{\alpha}(\omega)$ in Eq.~(\ref{lessg}) into Eq.~(\ref{rate1}) and writing the summation as an integral, we have
\begin{equation}
p_u = \int d\omega_{q} \sigma(\omega_{q})\frac{|a_{q}|^2N(3D)}{\hbar}\frac{2\Lambda_{q}}{((3D)^2-\omega_{q}^2)^2+\Lambda_{q}^2}.
\end{equation}
Since the denominator grows rapidly when $\omega_q$ goes away from $3D$, 
only a Lorentzian DOS peaking near the energy $\omega=3D$ can effectively contribute to the integral. For integration with respect to such a DOS function (i.e., set $\sigma(\omega_q)=\sigma_{\alpha}(\omega_q)$), the integrand is shown in Fig.~\ref{lifetime}, which is similar to that in Eq.~(\ref{integrand}). The result can be obtained as
\begin{equation}
p_u = \frac{|a_q|^2N(3D)}{\hbar}\frac{2\omega_{\alpha}(3D\Gamma_{\alpha}+\omega_{\alpha}\Lambda_{q})}{(3D\omega_{\alpha}\Delta)^2+(3D\Gamma_{\alpha}+\omega_{\alpha}\Lambda_{q})^2}.\label{anhar}
\end{equation}
Here, in order to have the analytical expression, we neglect variation of $|a_q|$ and $\Lambda_q$.

In $p_u$ on the above, $\Lambda_{q}$ (cf. Eq.~{\ref{broaden}}) and $N(3D)$ are functions of temperature. $N(3D)$ set a barrier of $3D$ and its multiplication with Bose-Einstein function $N(\omega_{\alpha}), N(\omega_{\beta})$ increases the barrier, so in order to reduce the barrier we should make the $\Lambda_{q}$ term in the denominator dominant. Namely, Eq.~(\ref{anhar}) should reduce to 
\begin{equation}
p_u(\omega)\propto\frac{N(\omega)}{\Lambda_{q}},
\label{reson}
\end{equation}
for which two conditions should be satisfied. First, $\Delta$ should be small, so that
\begin{equation}
p_u \simeq \frac{|a_q|^2N(3D)}{\hbar}\frac{2\omega_{\alpha}}{3D\Gamma_{\alpha}+\omega_{\alpha}\Lambda_{q}}.
\end{equation}
This requires that a vibrational mode lies around the transition energy $\omega=3D$, a resonance condition. Further, for the term $\omega_{\alpha}\Lambda_{q}$ to be dominant, it is required that $3D\Gamma_{\alpha}\ll \omega_{\alpha}\Lambda_{q}$, that is, the anharmonic interaction among the phonons should be much stronger than interaction between the Debye phonons and the vibrational mode $\alpha$.

\begin{figure}
\centering
\includegraphics[width=0.6\textwidth]{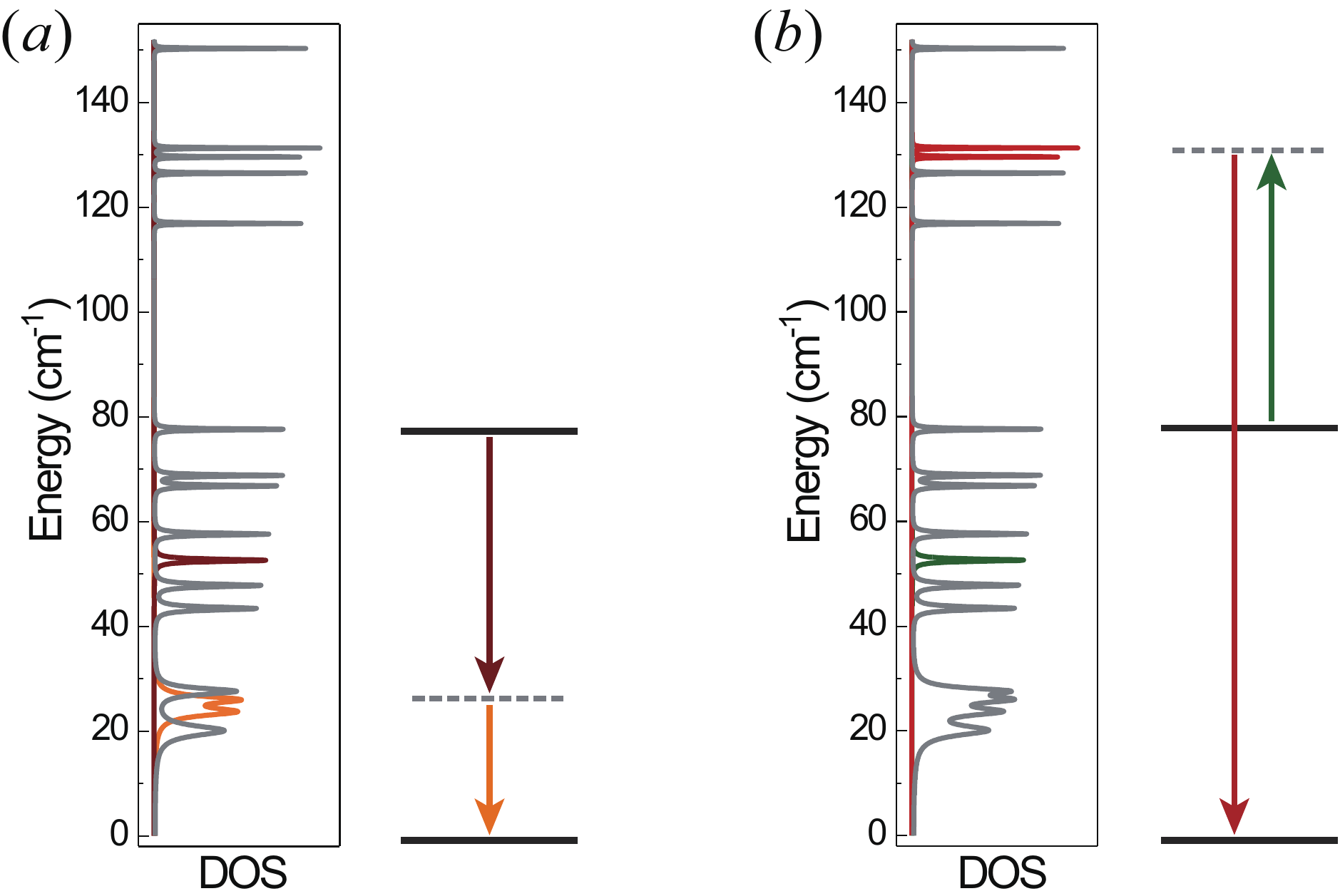}
\caption{Due to energy conservation and rapid DOS descending away from the peaks, only phonons around two vibrational modes satisfying $\omega_{\beta}\pm \omega_{\alpha}\approx 3D$ can effectively contribute to decay of the phonons of energy $\omega=3D$. (a) Double phonon decaying and (b) Ramman decaying are both supported by the vibration spectrum of [tpa$^{\ph}$Fe]$^{-1}$. The colored DOS peaks represent the phonons that contribute to the corresponding processes. According to Eq.~(\ref{broaden}), the Raman decaying is dominated over by the double phonon decaying, since $N(\omega_{\alpha}), N(\omega_{\beta})\ll 1$.}
\label{spectrum}
\end{figure}

Then we can see that another resonance condition is required to effectively reduce the barrier. If $\Lambda_{q}$ in Eq.~(\ref{broaden}) is dominated by one of the Bose-Einstein functions (say $N(\omega_{\alpha})$), the transition rate would take the form $p_u\approx N(\omega)/N(\omega_{\alpha})$, yielding $\tau\propto e^{(3D-\omega_\alpha)/k_BT}$. However, $N(\omega_{\alpha})\ll 1$ for typical vibrational modes and experimental temperatures. In general, the addend $1$ in the double phonon term (first term in Eq.~(\ref{broaden})) dominates over the Bose-Einstein functions, and resulting in no obvious barrier reduction. The reduction occurs, only if the condition for the Raman decaying is satisfied, and the double phonon decaying is suppressed. Namely, it is required that two of the vibrational modes have energies $\omega_{\alpha}-\omega_{\beta}\approx\omega$ and no mode pair satisfies $\omega_{\alpha}+\omega_{\beta}\approx\omega$. The [tpa$^{\ph}$Fe]$^{-1}$ example is not the case, since the vibration spectrum supports both decaying processes (Fig.~\ref{spectrum}). While this barrier reduction mechanism may account for some observed under barrier relaxation, it relies on restrictive conditions. Regarding prevalence of the under barrier relaxation, an explanation based on a more general mechanism is needed.

\subsection{Effect of the second order processes}
The second order spin-phonon coupling takes the form $H_2 = \sum_{qq'}\frac{\partial^2H_{spin}}{\partial V_{q}\partial V_{q'}}V_{q}V_{q'}$, i.e., $A_{qq'}=\frac{\partial^2H_{spin}}{\partial V_{q}\partial V_{q'}}$. Here, the pair $(q,q')$ should be taken as a single phononic degree of freedom. The correlation function now takes the form
\begin{equation}
\Gamma_{qq'}(\omega) = \frac{1}{\hbar^2} \int_{0}^{+\infty} \de e^{i\omega s}\langle V_{q'}^{\dagger}(t-s)V_{q}^{\dagger}(t-s)V_{q}(s)V_{q'}(s)\rangle.
\end{equation}
It is proper to define the double phonon Green's functions as
\begin{equation}
G_{qq'}(t,t')=-\frac{i}{\hbar}\langle V_{q}(t)V_{q'}(t) V_{q}^{\dagger}(t')V_{q'}^{\dagger}(t')\rangle,
\end{equation}
which is consistent with the main text Eq.~(1).
Since the displacement operators are commutable,
\begin{align}
&[V_q, V_{q'}]=0\\
&[V_q, V_{q'}^{\dagger}]=0
\end{align}
it is free to change their order and the Wick theorem gives
\begin{equation}
G_{qq'}^<(t,t')=i\hbar G_{q}^<(t,t')G_{q'}^<(t,t'),
\end{equation}
whose Fourier transform reads
\begin{equation}
G_{qq'}^<(\omega) = \int  \frac{i\hbar}{2\pi} d \omega'  G_{\alpha}^<(\omega)G_{\beta}^<(\omega-\omega').
\end{equation}
The single phonon Green's function is defined as
\begin{equation}
G_q(t,t') = -\frac{i}{\hbar}\langle V_q(t) V_q^{\dagger}(t')\rangle.
\end{equation}
In the definition of the Green's functions, $\langle\ \rangle$ means the contour-ordered average~\cite{Stefanucci2013}.

Substituting the single phonon lesser Green's functions with Eq.~(\ref{single}), we get an expression similar to Eq.~(\ref{sigma2}), differed by a factor. Accordingly, the upward transition rate (Eq.~(\ref{rate1})) is given by
\begin{align}
p_u =  N(\omega)\iint \frac{\pi|a_{qq'}|^2}{2\omega_{q}\omega_{q'}}\de\omega_{q}\de\omega_{q'}\sigma(\omega_{q})\sigma(\omega_{q'}) \{&[N(\omega_{q})+N(\omega_{q'})+1]\delta(\omega-\omega_{q}-\omega_{q'})\nonumber\\
+ &[N(\omega_{q})-N(\omega_{q'})]\delta(\omega+\omega_{q}-\omega_{q'})\},
\label{spu2}
\end{align}
with $\omega=3D$ specifying the energy gain. By energy conservation, we can identify the first term as the double phonon process whereby two phonons are absorbed, and the second terms as the Raman process whereby a phonon is absorbed ($\omega_{q'}$) and a phonon of lower energy is emitted ($\omega_{q}$). In this formulation, the detailed balance is very clear, since ratio between the rate of upward and downward transition is $N(\omega)/N(-\omega)=-e^{-\omega/k_BT}$ with $\omega>0$. We note that Eq.~(\ref{spu2}) is equivalent to the usual formulations of the Raman and double phonon process. By moving the factor $N(\omega)$ into the integral, the formulation can be factorized as productions $N(\omega_q)[1+N(\omega+\omega_q)]$ (Raman) and $N(\omega_q)N(\omega-\omega_q)$ (double phonon).

At high temperatures, the vibrational-mode-broadening resulted phonons are accessible. Due to $N(\omega_q), N(\omega_{q'})\ll 1$, in general the double phonon process is dominant. Carrying out the integration with respect to two Lorentzian DOS peaks satisfying $\omega_{\alpha}+\omega_{\beta}-\omega=\Delta$, we have
\begin{equation}
p_u \simeq \frac{2 |a_{\alpha\beta}|^2(\omega_{\alpha}\Gamma_{\beta}+\omega_{\beta}\Gamma_{\alpha})}{(\omega_{\alpha}\omega_{\beta}\Delta)^2+(\omega_{\alpha}\Gamma_{\beta}+\omega_{\beta}\Gamma_{\alpha})^2}N(\omega).
\end{equation}
from which
\begin{equation}
\tau = \tau_0 e^{3D/k_BT} = \frac{(\omega_{\alpha}\omega_{\beta}\Delta)^2+(\omega_{\alpha}\Gamma_{\beta}+\omega_{\beta}\Gamma_{\alpha})^2}{2 |a_{\alpha\beta}|^2(\omega_{\alpha}\Gamma_{\beta}+\omega_{\beta}\Gamma_{\alpha})}e^{3D/k_BT}.
\end{equation}
Here, $|a_{\alpha\beta}|$ approximates $|a_{qq'}|$ for the phonon pairs around the two DOS peaks. Thus, we have a $\tau_0$ dependent on broadening width and position of the mode pair in charge.

At low temperatures, as shown in the next section, it is likely that the direct tunneling between the ground state doublet dominates the relaxation. The result here is given for completeness. When temperature is reduced, $N(\omega_{\alpha}), N(\omega_{\beta})$ decay exponentially. So at certain temperature, the Raman and double phonon process involve a low energy Debye phonon (say, $\omega_q$) becomes dominant. The small $\omega_{q}$ expansion of $N(\omega_{q})$ leads to
\begin{align}
p_u = N(\omega)\iint \frac{\pi|a_{qq'}|^2}{2\omega_{q}\omega_{q'}}\de\omega_{q}\de\omega_{q'}\sigma(\omega_{q})\sigma(\omega_{q'})[(&\frac{k_BT}{\omega_q}+1)\delta(\omega-\omega_{q}-\omega_{q'})\nonumber\\
+ &\frac{k_BT}{\omega_q}\delta(\omega+\omega_{q}-\omega_{q'})].
\end{align}
Here, the $N(\omega_{q'})$ term is neglected, since it must be a vibrational-mode-broadening resulted phonons ($N(\omega_{q'})\ll N(\omega_{q})$). This is because the energy conversation $\omega_{q'}\pm \omega_{q}=3D$ implies $\omega_q'\gg \omega_D$, for $\omega_q < \omega_D$ and  $3D \gg \omega_D$ . In the temperature range $\omega_q \lesssim k_BT$, the above transition rate gives $\tau_0 \propto 1/T$. If one wishes, the integration can be analytically carried out by assuming the Debye DOS for $\omega_q$ and a Lorentzian DOS for $\omega_{q'}$.

\section{Direct tunneling}
\subsection{Effect of magnetic fields}
We first clarify some subtleties about the ground state doublet and relaxation caused by tunneling between the two states. When there is no external magnetic field, the ground state doublet of integer spins described by $H_{spin}=-DS_z^2-E(S_x^2-S_y^2)$ take the form $|S\rangle \pm |-S\rangle$ mixed with small portion of states $|S_z\neq \pm S\rangle$. For example, setting $D=26$ cm$^{-1}$, $E=0.005D$, the two lowest eigenstates of the $S=2$ molecule are given by
\begin{align}
|-\rangle&=0.70710(|-2\rangle+|2\rangle)+0.0044|0\rangle,\\
|+\rangle&=0.70711(|2\rangle-|-2\rangle),
\end{align}
with an energy difference $E_+-E_-=1.95\times10^{-3}$ cm$^{-1}$. For an initial polarization as the state $|2\rangle$ or $-2\rangle$, in the basis of $|-\rangle, |+\rangle$ the relaxation corresponds to diminishing of the off diagonal elements of the density operator, $\rho_{+-}$ and $\rho_{-+}$.

The situation significantly changes if a (even small) magnetic field is performed. For instance, under an magnetic field of $1500$ Oe~\cite{Harman2010}, the two lowest states become
\begin{align}
|-\rangle&=0.0015|2\rangle+0.0031|0\rangle+0.9999|-2\rangle,\\
|+\rangle&=0.9999|2\rangle+0.0031|0\rangle-0.0015|-2\rangle.
\end{align}
with an energy difference $E_+-E_-=0.64$ cm$^{-1}$. In this case, because of finite temperature and smallness of the splitting, an initial state being the ground state $|-\rangle$ is mixed with $|+\rangle$ by the relaxation. In the basis of $|-2\rangle, |2\rangle$, the relaxation corresponds to evolution $\rho_{-2,-2}\approx1.0\rightarrow\rho_{-2,-2}\approx0.5$ and $\rho_{2,2}\approx0.0\rightarrow\rho_{2,2}\approx0.5$. Since the relaxation-temperature-dependences are drawn from magnetic susceptibility measurements under a magnetic field, we adopt this setting, and the rate equation is given by Eq.~(2) in the main text.

\begin{figure}
\centering
\includegraphics[width=0.6\textwidth]{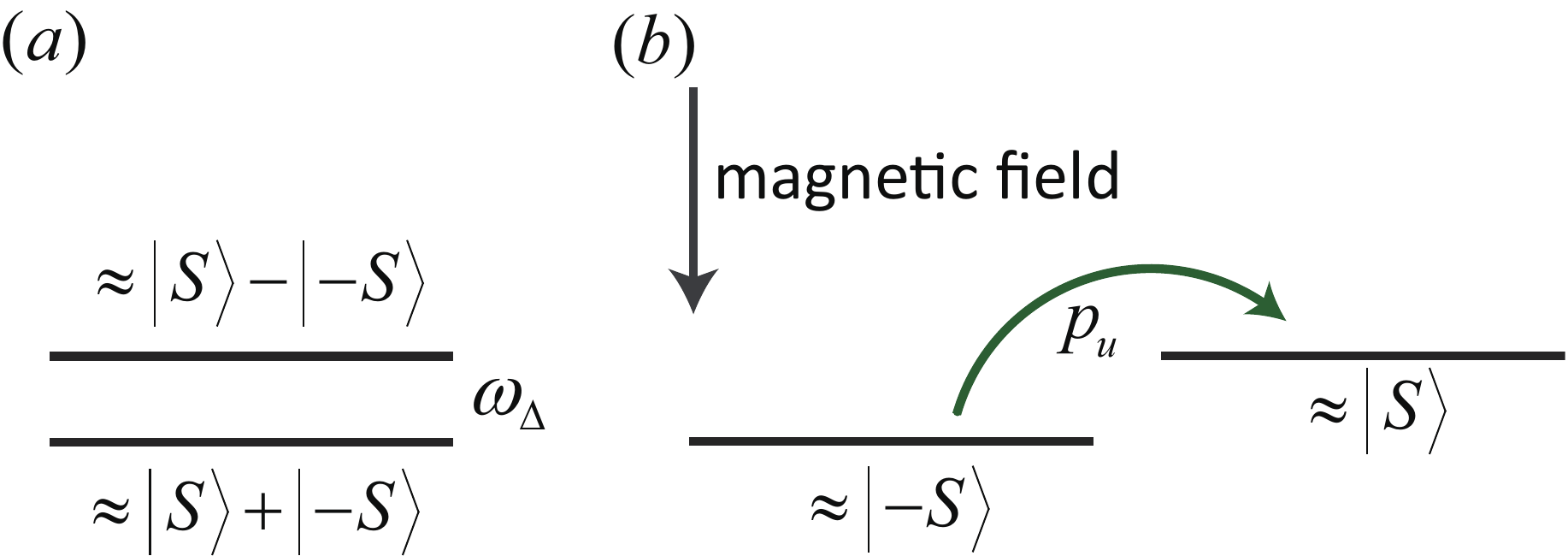}
\caption{(a) Without external magnetic filed, the transverse anisotropy leads to a pseudo ground state doublet that is mainly symmetric and antisymmetric combination of $|S\rangle, |-S\rangle$. When an external magnetic field is performed, the two lowest states are mainly consisted of $|S\rangle, |-S\rangle$, respectively. Since the latter situation is the usual experimental setting, we assume the relaxation is the transition as shown in (b).}
\label{tunneling}
\end{figure}

For half integer spins, if there are only even order magnetic anisotropies, only states with $\Delta S_z=\pm 2n$ ($n$ denotes positive integers) can be mixed to form eigenstates. States $|\pm S\rangle$ cannot be mixed, and the two lowest states are $|S\rangle+\sum_n c_n|S-2n\rangle$ and $|-S\rangle+\sum_n c'_n|-S+2n\rangle$ with small $c_n, c'_n$. Thus, the situation is similar to the case where a magnetic field is performed. If odd order magnetic anisotropies are present, states with $\Delta S_z=2n+1$ can also be mixed to form an eigenstate, and the above discussions about the effect of magnetic fields apply. Anyway, because the measurements are implemented under a magnetic field, we take the view that the relaxation process is the transition from the ground state to the other states that is slightly lifted by magnetic anisotropy and external magnetic fields (Fig.~\ref{tunneling}(b)).

\subsection{Exponential temperature dependence due to Raman process}
Result for the direct process is the same as Eq.~(\ref{direct}), except now $\omega=\omega_{\Delta}$ and an additional factor $2$, that is $\tau^{-1}=2\pi|a_q|^2\sigma(\omega_{\Delta})N(\omega_\Delta)/\hbar\omega_{\Delta}$. Here, we concentrate on derivation for the second order processes. As argued in the main text, the Raman process is dominant, and gives,
\begin{equation}
p_u = \pi N(\omega)\iint \frac{\de\omega_{q}\de\omega_{q'}}{\omega_{q}\omega_{q'}}|a_{qq'}|^2\sigma(\omega_{q})\sigma(\omega_{q'})[N(\omega_{q})-N(\omega_{q'})]\delta(\omega+\omega_{q}-\omega_{q'}).
\label{raman}
\end{equation}
Here, $\omega_{q'}=\omega_{q}+\omega_{\Delta}$ and smallness of $\omega_{\Delta}$ implies $\omega_{q'}\approx\omega_{q}$, that is, the two phonons lie around the same DOS peak (say, $\omega_{\alpha}$). However, if we approximate both $N(\omega_q), N(\omega_{q'})$ as $N(\omega_{\alpha})$, the result is zero. We need to take into account the difference $\omega_{\Delta}$. Since $\omega_q, \omega_{q'} \gg k_BT$ at low temperature, we have
\begin{align}
N(\omega_{q})-N(\omega_{q'})\simeq& e^{(-\omega_{q'}+\omega_{\Delta})/k_BT}-e^{-\omega_{q'}/k_BT}\\
=&(e^{\omega_{\Delta}/k_BT}-1)e^{-\omega_{q'}/k_BT}\\
=&\frac{e^{-\omega_{q'}/k_BT}}{N(\omega_{\Delta})}.
\end{align}
In Eq.~(\ref{raman}), $\omega=\omega_{\Delta}$, so the factor $N(\omega)$ is cancelled. The integration form is given by a similar integral as in Eq.~(\ref{integrand}), except now in Eq.~(\ref{raman}) the two DOS peaks are the same one, and $\Delta=\omega_{\Delta}$. Accordingly, the result is given by Eq.~(4) in the main text,
\begin{equation}
\tau^{-1}=2p_u\simeq\frac{4\omega_{\alpha} \Gamma_{\alpha}|a_{\alpha}|^2}{(\omega_{\alpha}^2\omega_{\Delta})^2+(2\omega_{\alpha}\Gamma_{\alpha})^2}e^{-\omega_{\alpha}/k_BT}.
\label{sresult}
\end{equation}

\subsection{Estimation of the parameters}
The calculation involves the second derivative of the transverse and high order magnetic anisotropies with respect to the atomic displacements. The smallness and sensitivity of magnetic anisotropy over atomic configuration pose a challenge to current ab initio calculation packages. As a work of general theory, here we do not seek to develop algorithms and codes, but give an estimation of the parameters, with which Fig.~3 in the main text is plotted.

Relaxation of the direct process reads $\tau^{-1}=2\pi|a_q|^2\sigma(\omega_{\Delta})N(\omega_\Delta)/\hbar\omega_{\Delta}$, and the temperature-independent part is
\begin{equation}
\tau_{01} = \frac{\hbar\omega_{\Delta}}{2\pi|a_q|^2\sigma(\omega_{\Delta})}.
\end{equation}
To have a concrete number for it, we use the values in Ref.~\cite{Harman2010}, $D=26$ cm$^{-1}$, $E=0.005D$, $g=2.28$, $H=1500$ Oe. These values lead to a splitting $\omega_{\Delta}\simeq 0.64$ cm$^{-1}$. The form of $\sigma(\omega_{\Delta})$ and value of $a_q$ are needed.

The spin-coupling strength is defined by $\partial H_{spin}/\partial V_q$. In practice, the derivatives are calculated from real space displacements. Due to the factor $1/\sqrt{N}$ (a mesh of $N$ $q$ points) in the Fourier transform of atomic displacements, the summation over $q$ does not depends on density of the $q$ mesh (as it should be). This fact amounts to normalized DOS functions in the integration form. The Lorentzian functions are naturally normalized to $1$. For the long wavelength Debye phonons, we have
\begin{equation}
\int_{0}^{\omega_D} \de \omega \frac{\omega^2}{C}= 1.
\label{DOSnorm}
\end{equation} 
which gives $C=\omega_D^3/3$, and $\sigma(\omega_{\Delta})=3\omega_{\Delta}^2/\omega_{D}^3$. It should be noted that the Debye phonons in this work refer to the phonons that well follow the Debye DOS function. Strictly speaking, they are the low energy part of the acoustic phonons. The right hand side of Eq.~(\ref{DOSnorm}) should be a number $<3$. Here, $1$ is used for order estimation.

\begin{figure}
\centering
\includegraphics[width=0.6\textwidth]{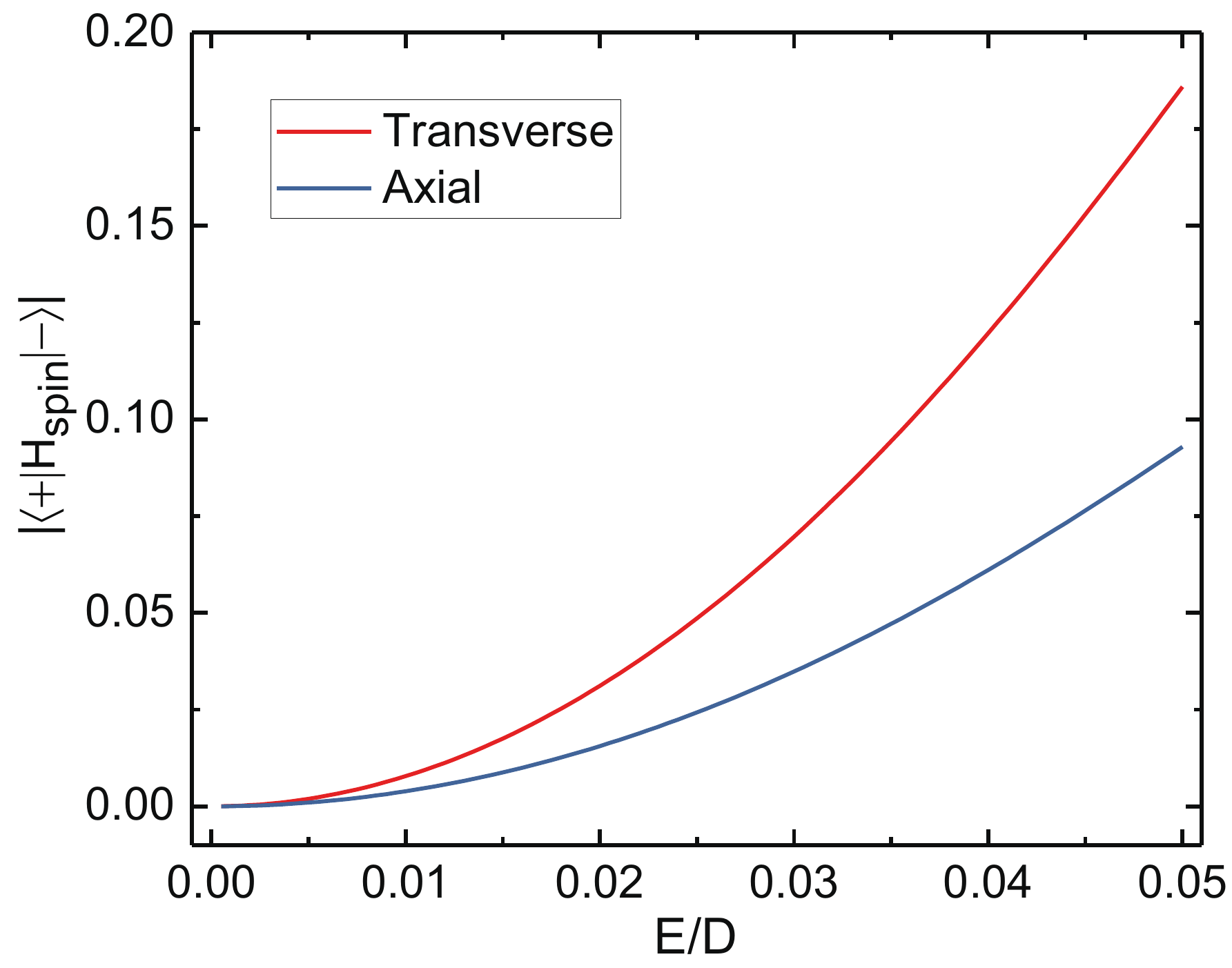}
\caption{Since magnitude of the transverse term $\langle+|E(S_x^2-S_y^2)|-\rangle$ is twice that of the axial term $\langle+|D S_z^2|-\rangle$, the transverse term is used for estimation of the transition operator element, i.e., $a_q\propto \langle+|E(S_x^2-S_y^2)|-\rangle$. The transition rate $p_u\propto |a_q|^2$ falls off, when $E/D$ is reduced. Thus, diminishing the transverse magnetic anisotropy is effective in lengthening the relaxation time.}
\label{rates}
\end{figure}

%Finite difference value of derivatives of the $\mathbf{D}$ matrix ($H_{spin}=\vec{S}\mathbf{D}\vec{S}$) $D$ are roughly in the order $\Delta D_{mn}/\Delta U_{j}\approx 0.01D_{mn}/\AA$~\cite{Lunghi}, where $U_j$ denotes atomic displacements. The vibrational modes are derived from the dynamical matrix defined as
%\begin{equation}
%\Psi_{jk} = \frac{\partial^2 P(\mathbf{U})}{\sqrt{m_j m_k} \partial U_j\partial U_k},
%\end{equation}
%where $P(\mathbf{U})$ is the dynamical potential and $m_j, m_k$ are atomic masses. It is convenient to define the mass normalized displacement $V_j=\sqrt{m_j}U_j$, and then the collective dynamical variable for mode $\alpha$ is given by
%\begin{equation}
%V_{\alpha} = \sum_j\epsilon^{\alpha}_j V_j,
%\end{equation}
%where $\epsilon^{\alpha}_j$ is the polarization vector of mode $\alpha$.

In Fig.~\ref{rates}, we plot magnitude of $\langle +|E(S_x^2-S_y^2)|-\rangle$ and $\langle +|D S_z^2|-\rangle$ for the example molecule under a magnetic field of $1500$ Oe. Besides reduction with decreasing $E/D$, value of the transverse magnetic anisotropy is twice that of the axial term.
For this reason, we use coupling between phonons and the transverse term $E(S_x^2-S_y^2)$ to estimate the tunneling rate.

Finite difference value of derivatives of the $\mathbf{D}$ matrix ($H_{spin}=\vec{S}\mathbf{D}\vec{S}$) are in the order $|\Delta D_{mn}|/\Delta U_{j}\approx 0.01|D_{mn}|/\AA$~\cite{Lunghi2017}, where $U_j$ denotes atomic displacements. The vibrational modes are derived from the dynamical matrix defined as
\begin{equation}
\Psi_{jk} = \frac{\partial^2 P(\mathbf{U})}{\sqrt{m_j m_k} \partial U_j\partial U_k},
\end{equation}
where $P(\mathbf{U})$ is the dynamical potential and $m_j, m_k$ are atomic masses. It is convenient to define the mass normalized displacement $V_j=\sqrt{m_j}U_j$, and then the collective dynamical variable for mode $\alpha$ is given by
\begin{equation}
V_{\alpha} = \sum_j\epsilon^{\alpha}_j V_j,
\end{equation}
where $\epsilon^{\alpha}_j$ is the polarization vector of mode $\alpha$. For phonons, the displacement $V_q$ is defined in a similar way.

We assume that the factor $0.01$ applies to the derivative of $E$, i.e., $|\Delta E|/\Delta U_{j}\approx 0.01|E|/\AA$. As shown in the main text, the intra-molecule motions for the Debye phonons are very small, and consequently so is the spin-phonon coupling. In this regard, we downscale the factor to $0.001$ for the Debye phonons, and estimate the matrix element of the transition operator as
\begin{equation}
a_{q1} = \frac{\Delta E}{\Delta V_q}\langle + | (S_x^2-S_y^2)|-\rangle=\frac{0.001E}{\sqrt{m}}\langle + | (S_x^2-S_y^2)|-\rangle,
\end{equation} 
where the atomic mass is added, since all our results assume mass normalized displacement. With $\omega_D=5$ cm$^{-1}$, a typical atom mass $m=50$ amu and $\omega_{\Delta}=0.64$ cm$^{-1}$, we have $\tau_{01}\approx 3900$ ns.

From Eq.~(\ref{sresult}), the temperature-independent part is
\begin{equation}
\tau_{02} = \frac{(\omega_{\alpha}^2\omega_{\Delta})^2+(2\omega_{\alpha}\Gamma_{\alpha})^2}{4\omega_{\alpha} \Gamma_{\alpha}|a_{\alpha}|^2}.
\end{equation}
For the second order spin phonon coupling, we assume that the vibrational-mode-broadening resulted phonons have similar coupling strength as the mode itself, and estimate the transition operator element as
\begin{equation}
a_{q2} = \frac{\Delta E}{(\Delta V_q)^2}\langle + | (S_x^2-S_y^2)|-\rangle=\frac{0.01E}{m}\langle + | (S_x^2+S_y^2)|-\rangle.
\end{equation}
Setting $\omega_{\alpha}=30$ cm$^{-1}$ and $\Gamma_{\alpha}=25$ cm$^{-2}$ (broadening of $5$ cm$^{-1}$), we have $\tau_{02}\approx 0.128$~ns, comparable with the experimental measurement~\cite{Harman2010}. The temperature range of the plottings in the main text Fig.~3 is $1/T=[0.06, 0.8]$ 1/K.

In Fig.~3(b) of the main text, energy of the higher vibrational mode is set as $30$ cm$^{-1}$ and the lower one is $10$ cm$^{-1}$. Ratio of the $\tau_{02}$ for the high ($\tau_{02}^h$) and low mode ($\tau_{02}^l$) is set as $\tau_{02}^h/\tau_{02}^l  
=100$. In principle, the presence of multiple barriers should be common, as it only requires that lower vibrational modes have weaker phonon-spin couplings. Reason for its less prevalence than the lower-than-expected barrier could be a resolution issue. Since the value of the barrier is fitted using data across a certain temperature span, it is likely that the value is an average of several modes. The double barrier phenomenon can clearly show up, only when two mode are distant enough and the lower mode has much smaller spin-phonon coupling. Otherwise, the lower mode would be dominant all over.

\bibliography{references}

\begin{thebibliography}{51}%
\makeatletter
\providecommand \@ifxundefined [1]{%
 \@ifx{#1\undefined}
}%
\providecommand \@ifnum [1]{%
 \ifnum #1\expandafter \@firstoftwo
 \else \expandafter \@secondoftwo
 \fi
}%
\providecommand \@ifx [1]{%
 \ifx #1\expandafter \@firstoftwo
 \else \expandafter \@secondoftwo
 \fi
}%
\providecommand \natexlab [1]{#1}%
\providecommand \enquote  [1]{``#1''}%
\providecommand \bibnamefont  [1]{#1}%
\providecommand \bibfnamefont [1]{#1}%
\providecommand \citenamefont [1]{#1}%
\providecommand \href@noop [0]{\@secondoftwo}%
\providecommand \href [0]{\begingroup \@sanitize@url \@href}%
\providecommand \@href[1]{\@@startlink{#1}\@@href}%
\providecommand \@@href[1]{\endgroup#1\@@endlink}%
\providecommand \@sanitize@url [0]{\catcode `\\12\catcode `\$12\catcode
  `\&12\catcode `\#12\catcode `\^12\catcode `\_12\catcode `\%12\relax}%
\providecommand \@@startlink[1]{}%
\providecommand \@@endlink[0]{}%
\providecommand \url  [0]{\begingroup\@sanitize@url \@url }%
\providecommand \@url [1]{\endgroup\@href {#1}{\urlprefix }}%
\providecommand \urlprefix  [0]{URL }%
\providecommand \Eprint [0]{\href }%
\providecommand \doibase [0]{http://dx.doi.org/}%
\providecommand \selectlanguage [0]{\@gobble}%
\providecommand \bibinfo  [0]{\@secondoftwo}%
\providecommand \bibfield  [0]{\@secondoftwo}%
\providecommand \translation [1]{[#1]}%
\providecommand \BibitemOpen [0]{}%
\providecommand \bibitemStop [0]{}%
\providecommand \bibitemNoStop [0]{.\EOS\space}%
\providecommand \EOS [0]{\spacefactor3000\relax}%
\providecommand \BibitemShut  [1]{\csname bibitem#1\endcsname}%
\let\auto@bib@innerbib\@empty
%</preamble>
\bibitem [{\citenamefont {DiVincenzo}(2000)}]{DiVincenzo2000}%
  \BibitemOpen
  \bibfield  {author} {\bibinfo {author} {\bibfnamefont {D.~P.}\ \bibnamefont
  {DiVincenzo}},\ }\href@noop {} {\bibfield  {journal} {\bibinfo  {journal}
  {Fortschritte Der Physik-Progress of Physics}\ }\textbf {\bibinfo {volume}
  {48}},\ \bibinfo {pages} {771} (\bibinfo {year} {2000})}\BibitemShut
  {NoStop}%
\bibitem [{\citenamefont {Escalera-Moreno}\ \emph
  {et~al.}(2018{\natexlab{a}})\citenamefont {Escalera-Moreno}, \citenamefont
  {Baldoví}, \citenamefont {Gaita-Ariño},\ and\ \citenamefont
  {Coronado}}]{Luis2018}%
  \BibitemOpen
  \bibfield  {author} {\bibinfo {author} {\bibfnamefont {L.}~\bibnamefont
  {Escalera-Moreno}}, \bibinfo {author} {\bibfnamefont {J.~J.}\ \bibnamefont
  {Baldoví}}, \bibinfo {author} {\bibfnamefont {A.}~\bibnamefont
  {Gaita-Ariño}}, \ and\ \bibinfo {author} {\bibfnamefont {E.}~\bibnamefont
  {Coronado}},\ }\href {\doibase 10.1039/C7SC05464E} {\bibfield  {journal}
  {\bibinfo  {journal} {Chem. Sci.}\ }\textbf {\bibinfo {volume} {9}},\
  \bibinfo {pages} {3265} (\bibinfo {year} {2018}{\natexlab{a}})}\BibitemShut
  {NoStop}%
\bibitem [{\citenamefont {Gaita-Ariño}\ \emph {et~al.}(2019)\citenamefont
  {Gaita-Ariño}, \citenamefont {Luis}, \citenamefont {Hill},\ and\
  \citenamefont {Coronado}}]{Gaita2019}%
  \BibitemOpen
  \bibfield  {author} {\bibinfo {author} {\bibfnamefont {A.}~\bibnamefont
  {Gaita-Ariño}}, \bibinfo {author} {\bibfnamefont {F.}~\bibnamefont {Luis}},
  \bibinfo {author} {\bibfnamefont {S.}~\bibnamefont {Hill}}, \ and\ \bibinfo
  {author} {\bibfnamefont {E.}~\bibnamefont {Coronado}},\ }\href {\doibase
  10.1038/s41557-019-0232-y} {\bibfield  {journal} {\bibinfo  {journal} {Nature
  Chemistry}\ }\textbf {\bibinfo {volume} {11}},\ \bibinfo {pages} {301}
  (\bibinfo {year} {2019})}\BibitemShut {NoStop}%
\bibitem [{\citenamefont {Zadrozny}\ \emph {et~al.}(2013)\citenamefont
  {Zadrozny}, \citenamefont {Xiao}, \citenamefont {Atanasov}, \citenamefont
  {Long}, \citenamefont {Grandjean}, \citenamefont {Neese},\ and\ \citenamefont
  {Long}}]{Zadrozny2013}%
  \BibitemOpen
  \bibfield  {author} {\bibinfo {author} {\bibfnamefont {J.~M.}\ \bibnamefont
  {Zadrozny}}, \bibinfo {author} {\bibfnamefont {D.~J.}\ \bibnamefont {Xiao}},
  \bibinfo {author} {\bibfnamefont {M.}~\bibnamefont {Atanasov}}, \bibinfo
  {author} {\bibfnamefont {G.~J.}\ \bibnamefont {Long}}, \bibinfo {author}
  {\bibfnamefont {F.}~\bibnamefont {Grandjean}}, \bibinfo {author}
  {\bibfnamefont {F.}~\bibnamefont {Neese}}, \ and\ \bibinfo {author}
  {\bibfnamefont {J.~R.}\ \bibnamefont {Long}},\ }\href {\doibase
  10.1038/nchem.1630} {\bibfield  {journal} {\bibinfo  {journal} {Nature
  Chemistry}\ }\textbf {\bibinfo {volume} {5}},\ \bibinfo {pages} {577–581}
  (\bibinfo {year} {2013})}\BibitemShut {NoStop}%
\bibitem [{\citenamefont {Blagg}\ \emph {et~al.}(2013)\citenamefont {Blagg},
  \citenamefont {Ungur}, \citenamefont {Tuna}, \citenamefont {Speak},
  \citenamefont {Comar}, \citenamefont {Collison}, \citenamefont {Wernsdorfer},
  \citenamefont {McInnes}, \citenamefont {Chibotaru},\ and\ \citenamefont
  {Winpenny}}]{Blagg2013}%
  \BibitemOpen
  \bibfield  {author} {\bibinfo {author} {\bibfnamefont {R.~J.}\ \bibnamefont
  {Blagg}}, \bibinfo {author} {\bibfnamefont {L.}~\bibnamefont {Ungur}},
  \bibinfo {author} {\bibfnamefont {F.}~\bibnamefont {Tuna}}, \bibinfo {author}
  {\bibfnamefont {J.}~\bibnamefont {Speak}}, \bibinfo {author} {\bibfnamefont
  {P.}~\bibnamefont {Comar}}, \bibinfo {author} {\bibfnamefont
  {D.}~\bibnamefont {Collison}}, \bibinfo {author} {\bibfnamefont
  {W.}~\bibnamefont {Wernsdorfer}}, \bibinfo {author} {\bibfnamefont
  {E.~J.~L.}\ \bibnamefont {McInnes}}, \bibinfo {author} {\bibfnamefont
  {L.~F.}\ \bibnamefont {Chibotaru}}, \ and\ \bibinfo {author} {\bibfnamefont
  {R.~E.~P.}\ \bibnamefont {Winpenny}},\ }\href {\doibase 10.1038/nchem.1707}
  {\bibfield  {journal} {\bibinfo  {journal} {Nature Chemistry}\ }\textbf
  {\bibinfo {volume} {5}},\ \bibinfo {pages} {673} (\bibinfo {year}
  {2013})}\BibitemShut {NoStop}%
\bibitem [{\citenamefont {Chen}\ \emph {et~al.}(2016)\citenamefont {Chen},
  \citenamefont {Liu}, \citenamefont {Ungur}, \citenamefont {Liu},
  \citenamefont {Li}, \citenamefont {Wang}, \citenamefont {Ni}, \citenamefont
  {Chibotaru}, \citenamefont {Chen},\ and\ \citenamefont {Tong}}]{Chen2016}%
  \BibitemOpen
  \bibfield  {author} {\bibinfo {author} {\bibfnamefont {Y.-C.}\ \bibnamefont
  {Chen}}, \bibinfo {author} {\bibfnamefont {J.-L.}\ \bibnamefont {Liu}},
  \bibinfo {author} {\bibfnamefont {L.}~\bibnamefont {Ungur}}, \bibinfo
  {author} {\bibfnamefont {J.}~\bibnamefont {Liu}}, \bibinfo {author}
  {\bibfnamefont {Q.-W.}\ \bibnamefont {Li}}, \bibinfo {author} {\bibfnamefont
  {L.-F.}\ \bibnamefont {Wang}}, \bibinfo {author} {\bibfnamefont {Z.-P.}\
  \bibnamefont {Ni}}, \bibinfo {author} {\bibfnamefont {L.~F.}\ \bibnamefont
  {Chibotaru}}, \bibinfo {author} {\bibfnamefont {X.-M.}\ \bibnamefont {Chen}},
  \ and\ \bibinfo {author} {\bibfnamefont {M.-L.}\ \bibnamefont {Tong}},\
  }\href {\doibase 10.1021/jacs.5b13584} {\bibfield  {journal} {\bibinfo
  {journal} {Journal of the American Chemical Society}\ }\textbf {\bibinfo
  {volume} {138}},\ \bibinfo {pages} {2829} (\bibinfo {year} {2016})},\
  \bibinfo {note} {pMID: 26883386}\BibitemShut {NoStop}%
\bibitem [{\citenamefont {Goodwin}\ \emph {et~al.}(2017)\citenamefont
  {Goodwin}, \citenamefont {Ortu}, \citenamefont {Reta}, \citenamefont
  {Chilton},\ and\ \citenamefont {Mills}}]{Goodwin2017}%
  \BibitemOpen
  \bibfield  {author} {\bibinfo {author} {\bibfnamefont {C.~A.~P.}\
  \bibnamefont {Goodwin}}, \bibinfo {author} {\bibfnamefont {F.}~\bibnamefont
  {Ortu}}, \bibinfo {author} {\bibfnamefont {D.}~\bibnamefont {Reta}}, \bibinfo
  {author} {\bibfnamefont {N.~F.}\ \bibnamefont {Chilton}}, \ and\ \bibinfo
  {author} {\bibfnamefont {D.~P.}\ \bibnamefont {Mills}},\ }\href {\doibase
  10.1038/nature23447} {\bibfield  {journal} {\bibinfo  {journal} {Nature}\
  }\textbf {\bibinfo {volume} {548}},\ \bibinfo {pages} {439} (\bibinfo {year}
  {2017})}\BibitemShut {NoStop}%
\bibitem [{\citenamefont {Guo}\ \emph {et~al.}(2018)\citenamefont {Guo},
  \citenamefont {Day}, \citenamefont {Chen}, \citenamefont {Tong},
  \citenamefont {Mansikkam{\"a}ki},\ and\ \citenamefont {Layfield}}]{Guo2018}%
  \BibitemOpen
  \bibfield  {author} {\bibinfo {author} {\bibfnamefont {F.-S.}\ \bibnamefont
  {Guo}}, \bibinfo {author} {\bibfnamefont {B.~M.}\ \bibnamefont {Day}},
  \bibinfo {author} {\bibfnamefont {Y.-C.}\ \bibnamefont {Chen}}, \bibinfo
  {author} {\bibfnamefont {M.-L.}\ \bibnamefont {Tong}}, \bibinfo {author}
  {\bibfnamefont {A.}~\bibnamefont {Mansikkam{\"a}ki}}, \ and\ \bibinfo
  {author} {\bibfnamefont {R.~A.}\ \bibnamefont {Layfield}},\ }\href {\doibase
  10.1126/science.aav0652} {\bibfield  {journal} {\bibinfo  {journal}
  {Science}\ }\textbf {\bibinfo {volume} {362}},\ \bibinfo {pages} {1400}
  (\bibinfo {year} {2018})}\BibitemShut {NoStop}%
\bibitem [{\citenamefont {Randall~McClain}\ \emph {et~al.}(2018)\citenamefont
  {Randall~McClain}, \citenamefont {Gould}, \citenamefont {Chakarawet},
  \citenamefont {Teat}, \citenamefont {Groshens}, \citenamefont {Long},\ and\
  \citenamefont {Harvey}}]{Randall2018}%
  \BibitemOpen
  \bibfield  {author} {\bibinfo {author} {\bibfnamefont {K.}~\bibnamefont
  {Randall~McClain}}, \bibinfo {author} {\bibfnamefont {C.~A.}\ \bibnamefont
  {Gould}}, \bibinfo {author} {\bibfnamefont {K.}~\bibnamefont {Chakarawet}},
  \bibinfo {author} {\bibfnamefont {S.~J.}\ \bibnamefont {Teat}}, \bibinfo
  {author} {\bibfnamefont {T.~J.}\ \bibnamefont {Groshens}}, \bibinfo {author}
  {\bibfnamefont {J.~R.}\ \bibnamefont {Long}}, \ and\ \bibinfo {author}
  {\bibfnamefont {B.~G.}\ \bibnamefont {Harvey}},\ }\href {\doibase
  10.1039/C8SC03907K} {\bibfield  {journal} {\bibinfo  {journal} {Chem. Sci.}\
  }\textbf {\bibinfo {volume} {9}},\ \bibinfo {pages} {8492} (\bibinfo {year}
  {2018})}\BibitemShut {NoStop}%
\bibitem [{\citenamefont {Ishikawa}\ \emph {et~al.}(2003)\citenamefont
  {Ishikawa}, \citenamefont {Sugita}, \citenamefont {Ishikawa}, \citenamefont
  {Koshihara},\ and\ \citenamefont {Kaizu}}]{Ishikawa2003}%
  \BibitemOpen
  \bibfield  {author} {\bibinfo {author} {\bibfnamefont {N.}~\bibnamefont
  {Ishikawa}}, \bibinfo {author} {\bibfnamefont {M.}~\bibnamefont {Sugita}},
  \bibinfo {author} {\bibfnamefont {T.}~\bibnamefont {Ishikawa}}, \bibinfo
  {author} {\bibfnamefont {S.-y.}\ \bibnamefont {Koshihara}}, \ and\ \bibinfo
  {author} {\bibfnamefont {Y.}~\bibnamefont {Kaizu}},\ }\href {\doibase
  10.1021/ja029629n} {\bibfield  {journal} {\bibinfo  {journal} {Journal of the
  American Chemical Society}\ }\textbf {\bibinfo {volume} {125}},\ \bibinfo
  {pages} {8694} (\bibinfo {year} {2003})},\ \bibinfo {note} {pMID: 12862442},\
  \Eprint {http://arxiv.org/abs/https://doi.org/10.1021/ja029629n}
  {https://doi.org/10.1021/ja029629n} \BibitemShut {NoStop}%
\bibitem [{\citenamefont {Ardavan}\ \emph {et~al.}(2007)\citenamefont
  {Ardavan}, \citenamefont {Rival}, \citenamefont {Morton}, \citenamefont
  {Blundell}, \citenamefont {Tyryshkin}, \citenamefont {Timco},\ and\
  \citenamefont {Winpenny}}]{Ardavan2007}%
  \BibitemOpen
  \bibfield  {author} {\bibinfo {author} {\bibfnamefont {A.}~\bibnamefont
  {Ardavan}}, \bibinfo {author} {\bibfnamefont {O.}~\bibnamefont {Rival}},
  \bibinfo {author} {\bibfnamefont {J.~J.~L.}\ \bibnamefont {Morton}}, \bibinfo
  {author} {\bibfnamefont {S.~J.}\ \bibnamefont {Blundell}}, \bibinfo {author}
  {\bibfnamefont {A.~M.}\ \bibnamefont {Tyryshkin}}, \bibinfo {author}
  {\bibfnamefont {G.~A.}\ \bibnamefont {Timco}}, \ and\ \bibinfo {author}
  {\bibfnamefont {R.~E.~P.}\ \bibnamefont {Winpenny}},\ }\href {\doibase
  10.1103/PhysRevLett.98.057201} {\bibfield  {journal} {\bibinfo  {journal}
  {Phys. Rev. Lett.}\ }\textbf {\bibinfo {volume} {98}},\ \bibinfo {pages}
  {057201} (\bibinfo {year} {2007})}\BibitemShut {NoStop}%
\bibitem [{\citenamefont {Magnani}\ \emph {et~al.}(2010)\citenamefont
  {Magnani}, \citenamefont {Colineau}, \citenamefont {Eloirdi}, \citenamefont
  {Griveau}, \citenamefont {Caciuffo}, \citenamefont {Cornet}, \citenamefont
  {May}, \citenamefont {Sharrad}, \citenamefont {Collison},\ and\ \citenamefont
  {Winpenny}}]{Magnani2010}%
  \BibitemOpen
  \bibfield  {author} {\bibinfo {author} {\bibfnamefont {N.}~\bibnamefont
  {Magnani}}, \bibinfo {author} {\bibfnamefont {E.}~\bibnamefont {Colineau}},
  \bibinfo {author} {\bibfnamefont {R.}~\bibnamefont {Eloirdi}}, \bibinfo
  {author} {\bibfnamefont {J.-C.}\ \bibnamefont {Griveau}}, \bibinfo {author}
  {\bibfnamefont {R.}~\bibnamefont {Caciuffo}}, \bibinfo {author}
  {\bibfnamefont {S.~M.}\ \bibnamefont {Cornet}}, \bibinfo {author}
  {\bibfnamefont {I.}~\bibnamefont {May}}, \bibinfo {author} {\bibfnamefont
  {C.~A.}\ \bibnamefont {Sharrad}}, \bibinfo {author} {\bibfnamefont
  {D.}~\bibnamefont {Collison}}, \ and\ \bibinfo {author} {\bibfnamefont
  {R.~E.~P.}\ \bibnamefont {Winpenny}},\ }\href {\doibase
  10.1103/PhysRevLett.104.197202} {\bibfield  {journal} {\bibinfo  {journal}
  {Phys. Rev. Lett.}\ }\textbf {\bibinfo {volume} {104}},\ \bibinfo {pages}
  {197202} (\bibinfo {year} {2010})}\BibitemShut {NoStop}%
\bibitem [{\citenamefont {Harman}\ \emph {et~al.}(2010)\citenamefont {Harman},
  \citenamefont {Harris}, \citenamefont {Freedman}, \citenamefont {Fong},
  \citenamefont {Chang}, \citenamefont {Rinehart}, \citenamefont {Ozarowski},
  \citenamefont {Sougrati}, \citenamefont {Grandjean}, \citenamefont {Long},
  \citenamefont {Long},\ and\ \citenamefont {Chang}}]{Harman2010}%
  \BibitemOpen
  \bibfield  {author} {\bibinfo {author} {\bibfnamefont {W.~H.}\ \bibnamefont
  {Harman}}, \bibinfo {author} {\bibfnamefont {T.~D.}\ \bibnamefont {Harris}},
  \bibinfo {author} {\bibfnamefont {D.~E.}\ \bibnamefont {Freedman}}, \bibinfo
  {author} {\bibfnamefont {H.}~\bibnamefont {Fong}}, \bibinfo {author}
  {\bibfnamefont {A.}~\bibnamefont {Chang}}, \bibinfo {author} {\bibfnamefont
  {J.~D.}\ \bibnamefont {Rinehart}}, \bibinfo {author} {\bibfnamefont
  {A.}~\bibnamefont {Ozarowski}}, \bibinfo {author} {\bibfnamefont {M.~T.}\
  \bibnamefont {Sougrati}}, \bibinfo {author} {\bibfnamefont {F.}~\bibnamefont
  {Grandjean}}, \bibinfo {author} {\bibfnamefont {G.~J.}\ \bibnamefont {Long}},
  \bibinfo {author} {\bibfnamefont {J.~R.}\ \bibnamefont {Long}}, \ and\
  \bibinfo {author} {\bibfnamefont {C.~J.}\ \bibnamefont {Chang}},\ }\href
  {\doibase 10.1021/ja105291x} {\bibfield  {journal} {\bibinfo  {journal}
  {Journal of the American Chemical Society}\ }\textbf {\bibinfo {volume}
  {132}},\ \bibinfo {pages} {18115} (\bibinfo {year} {2010})},\ \bibinfo {note}
  {pMID: 21141856}\BibitemShut {NoStop}%
\bibitem [{\citenamefont {Freedman}\ \emph {et~al.}(2010)\citenamefont
  {Freedman}, \citenamefont {Harman}, \citenamefont {Harris}, \citenamefont
  {Long}, \citenamefont {Chang},\ and\ \citenamefont {Long}}]{Freedman2010}%
  \BibitemOpen
  \bibfield  {author} {\bibinfo {author} {\bibfnamefont {D.~E.}\ \bibnamefont
  {Freedman}}, \bibinfo {author} {\bibfnamefont {W.~H.}\ \bibnamefont
  {Harman}}, \bibinfo {author} {\bibfnamefont {T.~D.}\ \bibnamefont {Harris}},
  \bibinfo {author} {\bibfnamefont {G.~J.}\ \bibnamefont {Long}}, \bibinfo
  {author} {\bibfnamefont {C.~J.}\ \bibnamefont {Chang}}, \ and\ \bibinfo
  {author} {\bibfnamefont {J.~R.}\ \bibnamefont {Long}},\ }\href {\doibase
  10.1021/ja909560d} {\bibfield  {journal} {\bibinfo  {journal} {Journal of the
  American Chemical Society}\ }\textbf {\bibinfo {volume} {132}},\ \bibinfo
  {pages} {1224} (\bibinfo {year} {2010})},\ \bibinfo {note} {pMID:
  20055389}\BibitemShut {NoStop}%
\bibitem [{\citenamefont {Zadrozny}\ and\ \citenamefont
  {Long}(2011)}]{Zadrozny2011}%
  \BibitemOpen
  \bibfield  {author} {\bibinfo {author} {\bibfnamefont {J.~M.}\ \bibnamefont
  {Zadrozny}}\ and\ \bibinfo {author} {\bibfnamefont {J.~R.}\ \bibnamefont
  {Long}},\ }\href {\doibase 10.1021/ja2100142} {\bibfield  {journal} {\bibinfo
   {journal} {Journal of the American Chemical Society}\ }\textbf {\bibinfo
  {volume} {133}},\ \bibinfo {pages} {20732} (\bibinfo {year} {2011})},\
  \bibinfo {note} {pMID: 22142241}\BibitemShut {NoStop}%
\bibitem [{\citenamefont {Lucaccini}\ \emph {et~al.}(2014)\citenamefont
  {Lucaccini}, \citenamefont {Sorace}, \citenamefont {Perfetti}, \citenamefont
  {Costes},\ and\ \citenamefont {Sessoli}}]{Lucaccini2014}%
  \BibitemOpen
  \bibfield  {author} {\bibinfo {author} {\bibfnamefont {E.}~\bibnamefont
  {Lucaccini}}, \bibinfo {author} {\bibfnamefont {L.}~\bibnamefont {Sorace}},
  \bibinfo {author} {\bibfnamefont {M.}~\bibnamefont {Perfetti}}, \bibinfo
  {author} {\bibfnamefont {J.-P.}\ \bibnamefont {Costes}}, \ and\ \bibinfo
  {author} {\bibfnamefont {R.}~\bibnamefont {Sessoli}},\ }\href {\doibase
  10.1039/C3CC48866G} {\bibfield  {journal} {\bibinfo  {journal} {Chem.
  Commun.}\ }\textbf {\bibinfo {volume} {50}},\ \bibinfo {pages} {1648}
  (\bibinfo {year} {2014})}\BibitemShut {NoStop}%
\bibitem [{\citenamefont {G{\'o}mez-Coca}\ \emph {et~al.}(2014)\citenamefont
  {G{\'o}mez-Coca}, \citenamefont {Urtizberea}, \citenamefont {Cremades},
  \citenamefont {Alonso}, \citenamefont {Cam{\'o}n}, \citenamefont {Ruiz},\
  and\ \citenamefont {Luis}}]{Gomez2014}%
  \BibitemOpen
  \bibfield  {author} {\bibinfo {author} {\bibfnamefont {S.}~\bibnamefont
  {G{\'o}mez-Coca}}, \bibinfo {author} {\bibfnamefont {A.}~\bibnamefont
  {Urtizberea}}, \bibinfo {author} {\bibfnamefont {E.}~\bibnamefont
  {Cremades}}, \bibinfo {author} {\bibfnamefont {P.~J.}\ \bibnamefont
  {Alonso}}, \bibinfo {author} {\bibfnamefont {A.}~\bibnamefont {Cam{\'o}n}},
  \bibinfo {author} {\bibfnamefont {E.}~\bibnamefont {Ruiz}}, \ and\ \bibinfo
  {author} {\bibfnamefont {F.}~\bibnamefont {Luis}},\ }\href {\doibase
  10.1038/ncomms5300} {\bibfield  {journal} {\bibinfo  {journal} {Nature
  Communications}\ }\textbf {\bibinfo {volume} {5}},\ \bibinfo {pages} {4300}
  (\bibinfo {year} {2014})}\BibitemShut {NoStop}%
\bibitem [{\citenamefont {Moseley}\ \emph {et~al.}(2018)\citenamefont
  {Moseley}, \citenamefont {Stavretis}, \citenamefont {Thirunavukkuarasu},
  \citenamefont {Ozerov}, \citenamefont {Cheng}, \citenamefont {Daemen},
  \citenamefont {Ludwig}, \citenamefont {Lu}, \citenamefont {Smirnov},
  \citenamefont {Brown}, \citenamefont {Pandey}, \citenamefont
  {Ramirez-Cuesta}, \citenamefont {Lamb}, \citenamefont {Atanasov},
  \citenamefont {Bill}, \citenamefont {Neese},\ and\ \citenamefont
  {Xue}}]{Moseley2018}%
  \BibitemOpen
  \bibfield  {author} {\bibinfo {author} {\bibfnamefont {D.~H.}\ \bibnamefont
  {Moseley}}, \bibinfo {author} {\bibfnamefont {S.~E.}\ \bibnamefont
  {Stavretis}}, \bibinfo {author} {\bibfnamefont {K.}~\bibnamefont
  {Thirunavukkuarasu}}, \bibinfo {author} {\bibfnamefont {M.}~\bibnamefont
  {Ozerov}}, \bibinfo {author} {\bibfnamefont {Y.}~\bibnamefont {Cheng}},
  \bibinfo {author} {\bibfnamefont {L.~L.}\ \bibnamefont {Daemen}}, \bibinfo
  {author} {\bibfnamefont {J.}~\bibnamefont {Ludwig}}, \bibinfo {author}
  {\bibfnamefont {Z.}~\bibnamefont {Lu}}, \bibinfo {author} {\bibfnamefont
  {D.}~\bibnamefont {Smirnov}}, \bibinfo {author} {\bibfnamefont {C.~M.}\
  \bibnamefont {Brown}}, \bibinfo {author} {\bibfnamefont {A.}~\bibnamefont
  {Pandey}}, \bibinfo {author} {\bibfnamefont {A.~J.}\ \bibnamefont
  {Ramirez-Cuesta}}, \bibinfo {author} {\bibfnamefont {A.~C.}\ \bibnamefont
  {Lamb}}, \bibinfo {author} {\bibfnamefont {M.}~\bibnamefont {Atanasov}},
  \bibinfo {author} {\bibfnamefont {E.}~\bibnamefont {Bill}}, \bibinfo {author}
  {\bibfnamefont {F.}~\bibnamefont {Neese}}, \ and\ \bibinfo {author}
  {\bibfnamefont {Z.-L.}\ \bibnamefont {Xue}},\ }\href {\doibase
  10.1038/s41467-018-04896-0} {\bibfield  {journal} {\bibinfo  {journal}
  {Nature Communications}\ }\textbf {\bibinfo {volume} {9}},\ \bibinfo {pages}
  {2572} (\bibinfo {year} {2018})}\BibitemShut {NoStop}%
\bibitem [{\citenamefont {Rajn{\'a}k}\ \emph {et~al.}(2019)\citenamefont
  {Rajn{\'a}k}, \citenamefont {Titi{\u s}}, \citenamefont {Moncoľ},
  \citenamefont {Renz},\ and\ \citenamefont {Bo{\u c}a}}]{Rajnak2019}%
  \BibitemOpen
  \bibfield  {author} {\bibinfo {author} {\bibfnamefont {C.}~\bibnamefont
  {Rajn{\'a}k}}, \bibinfo {author} {\bibfnamefont {J.}~\bibnamefont {Titi{\u
  s}}}, \bibinfo {author} {\bibfnamefont {J.}~\bibnamefont {Moncoľ}}, \bibinfo
  {author} {\bibfnamefont {F.}~\bibnamefont {Renz}}, \ and\ \bibinfo {author}
  {\bibfnamefont {R.}~\bibnamefont {Bo{\u c}a}},\ }\href {\doibase
  10.1039/C9CC06610A} {\bibfield  {journal} {\bibinfo  {journal} {Chem.
  Commun.}\ }\textbf {\bibinfo {volume} {55}},\ \bibinfo {pages} {13868}
  (\bibinfo {year} {2019})}\BibitemShut {NoStop}%
\bibitem [{\citenamefont {Garcia-Fernandez}\ \emph {et~al.}(2006)\citenamefont
  {Garcia-Fernandez}, \citenamefont {Bersuker},\ and\ \citenamefont
  {Boggs}}]{Garcia2006}%
  \BibitemOpen
  \bibfield  {author} {\bibinfo {author} {\bibfnamefont {P.}~\bibnamefont
  {Garcia-Fernandez}}, \bibinfo {author} {\bibfnamefont {I.~B.}\ \bibnamefont
  {Bersuker}}, \ and\ \bibinfo {author} {\bibfnamefont {J.~E.}\ \bibnamefont
  {Boggs}},\ }\href {\doibase 10.1063/1.2346682} {\bibfield  {journal}
  {\bibinfo  {journal} {The Journal of Chemical Physics}\ }\textbf {\bibinfo
  {volume} {125}},\ \bibinfo {pages} {104102} (\bibinfo {year} {2006})},\
  \Eprint {http://arxiv.org/abs/https://doi.org/10.1063/1.2346682}
  {https://doi.org/10.1063/1.2346682} \BibitemShut {NoStop}%
\bibitem [{\citenamefont {Garc\'{\i}a-Fern\'andez}\ \emph
  {et~al.}(2005)\citenamefont {Garc\'{\i}a-Fern\'andez}, \citenamefont
  {Bersuker}, \citenamefont {Aramburu}, \citenamefont {Barriuso},\ and\
  \citenamefont {Moreno}}]{Garcia2005}%
  \BibitemOpen
  \bibfield  {author} {\bibinfo {author} {\bibfnamefont {P.}~\bibnamefont
  {Garc\'{\i}a-Fern\'andez}}, \bibinfo {author} {\bibfnamefont {I.~B.}\
  \bibnamefont {Bersuker}}, \bibinfo {author} {\bibfnamefont {J.~A.}\
  \bibnamefont {Aramburu}}, \bibinfo {author} {\bibfnamefont {M.~T.}\
  \bibnamefont {Barriuso}}, \ and\ \bibinfo {author} {\bibfnamefont
  {M.}~\bibnamefont {Moreno}},\ }\href {\doibase 10.1103/PhysRevB.71.184117}
  {\bibfield  {journal} {\bibinfo  {journal} {Phys. Rev. B}\ }\textbf {\bibinfo
  {volume} {71}},\ \bibinfo {pages} {184117} (\bibinfo {year}
  {2005})}\BibitemShut {NoStop}%
\bibitem [{\citenamefont {Pae}\ and\ \citenamefont
  {Hizhnyakov}(2013)}]{Pae2013}%
  \BibitemOpen
  \bibfield  {author} {\bibinfo {author} {\bibfnamefont {K.}~\bibnamefont
  {Pae}}\ and\ \bibinfo {author} {\bibfnamefont {V.}~\bibnamefont
  {Hizhnyakov}},\ }\href {\doibase 10.1063/1.4792835} {\bibfield  {journal}
  {\bibinfo  {journal} {The Journal of Chemical Physics}\ }\textbf {\bibinfo
  {volume} {138}},\ \bibinfo {pages} {104103} (\bibinfo {year} {2013})},\
  \Eprint {http://arxiv.org/abs/https://doi.org/10.1063/1.4792835}
  {https://doi.org/10.1063/1.4792835} \BibitemShut {NoStop}%
\bibitem [{\citenamefont {Palii}\ \emph {et~al.}(2015)\citenamefont {Palii},
  \citenamefont {Ostrovsky}, \citenamefont {Reu}, \citenamefont {Tsukerblat},
  \citenamefont {Decurtins}, \citenamefont {Liu},\ and\ \citenamefont
  {Klokishner}}]{Palii2015}%
  \BibitemOpen
  \bibfield  {author} {\bibinfo {author} {\bibfnamefont {A.}~\bibnamefont
  {Palii}}, \bibinfo {author} {\bibfnamefont {S.}~\bibnamefont {Ostrovsky}},
  \bibinfo {author} {\bibfnamefont {O.}~\bibnamefont {Reu}}, \bibinfo {author}
  {\bibfnamefont {B.}~\bibnamefont {Tsukerblat}}, \bibinfo {author}
  {\bibfnamefont {S.}~\bibnamefont {Decurtins}}, \bibinfo {author}
  {\bibfnamefont {S.-X.}\ \bibnamefont {Liu}}, \ and\ \bibinfo {author}
  {\bibfnamefont {S.}~\bibnamefont {Klokishner}},\ }\href {\doibase
  10.1063/1.4928642} {\bibfield  {journal} {\bibinfo  {journal} {The Journal of
  Chemical Physics}\ }\textbf {\bibinfo {volume} {143}},\ \bibinfo {pages}
  {084502} (\bibinfo {year} {2015})},\ \Eprint
  {http://arxiv.org/abs/https://doi.org/10.1063/1.4928642}
  {https://doi.org/10.1063/1.4928642} \BibitemShut {NoStop}%
\bibitem [{\citenamefont {Escalera-Moreno}\ \emph {et~al.}(2017)\citenamefont
  {Escalera-Moreno}, \citenamefont {Suaud}, \citenamefont {Gaita-Ariño},\ and\
  \citenamefont {Coronado}}]{Moreno2017}%
  \BibitemOpen
  \bibfield  {author} {\bibinfo {author} {\bibfnamefont {L.}~\bibnamefont
  {Escalera-Moreno}}, \bibinfo {author} {\bibfnamefont {N.}~\bibnamefont
  {Suaud}}, \bibinfo {author} {\bibfnamefont {A.}~\bibnamefont {Gaita-Ariño}},
  \ and\ \bibinfo {author} {\bibfnamefont {E.}~\bibnamefont {Coronado}},\
  }\href {\doibase 10.1021/acs.jpclett.7b00479} {\bibfield  {journal} {\bibinfo
   {journal} {The Journal of Physical Chemistry Letters}\ }\textbf {\bibinfo
  {volume} {8}},\ \bibinfo {pages} {1695} (\bibinfo {year} {2017})},\ \bibinfo
  {note} {pMID: 28350165},\ \Eprint
  {http://arxiv.org/abs/https://doi.org/10.1021/acs.jpclett.7b00479}
  {https://doi.org/10.1021/acs.jpclett.7b00479} \BibitemShut {NoStop}%
\bibitem [{\citenamefont {Lunghi}\ \emph {et~al.}(2017)\citenamefont {Lunghi},
  \citenamefont {Totti}, \citenamefont {Sessoli},\ and\ \citenamefont
  {Sanvito}}]{Lunghi2017}%
  \BibitemOpen
  \bibfield  {author} {\bibinfo {author} {\bibfnamefont {A.}~\bibnamefont
  {Lunghi}}, \bibinfo {author} {\bibfnamefont {F.}~\bibnamefont {Totti}},
  \bibinfo {author} {\bibfnamefont {R.}~\bibnamefont {Sessoli}}, \ and\
  \bibinfo {author} {\bibfnamefont {S.}~\bibnamefont {Sanvito}},\ }\href
  {\doibase 10.1038/ncomms14620} {\bibfield  {journal} {\bibinfo  {journal}
  {Nature Communications}\ }\textbf {\bibinfo {volume} {8}},\ \bibinfo {pages}
  {14620} (\bibinfo {year} {2017})}\BibitemShut {NoStop}%
\bibitem [{\citenamefont {Shrivastava}(1983)}]{Shrivastava1983}%
  \BibitemOpen
  \bibfield  {author} {\bibinfo {author} {\bibfnamefont {K.~N.}\ \bibnamefont
  {Shrivastava}},\ }\href@noop {} {\bibfield  {journal} {\bibinfo  {journal}
  {phys. stat. sol, (b)}\ }\textbf {\bibinfo {volume} {117}},\ \bibinfo {pages}
  {437} (\bibinfo {year} {1983})}\BibitemShut {NoStop}%
\bibitem [{\citenamefont {Giraud}\ \emph {et~al.}(2001)\citenamefont {Giraud},
  \citenamefont {Wernsdorfer}, \citenamefont {Tkachuk}, \citenamefont
  {Mailly},\ and\ \citenamefont {Barbara}}]{Giraud2001}%
  \BibitemOpen
  \bibfield  {author} {\bibinfo {author} {\bibfnamefont {R.}~\bibnamefont
  {Giraud}}, \bibinfo {author} {\bibfnamefont {W.}~\bibnamefont {Wernsdorfer}},
  \bibinfo {author} {\bibfnamefont {A.~M.}\ \bibnamefont {Tkachuk}}, \bibinfo
  {author} {\bibfnamefont {D.}~\bibnamefont {Mailly}}, \ and\ \bibinfo {author}
  {\bibfnamefont {B.}~\bibnamefont {Barbara}},\ }\href {\doibase
  10.1103/PhysRevLett.87.057203} {\bibfield  {journal} {\bibinfo  {journal}
  {Phys. Rev. Lett.}\ }\textbf {\bibinfo {volume} {87}},\ \bibinfo {pages}
  {057203} (\bibinfo {year} {2001})}\BibitemShut {NoStop}%
\bibitem [{\citenamefont {Ishikawa}\ \emph {et~al.}(2005)\citenamefont
  {Ishikawa}, \citenamefont {Sugita},\ and\ \citenamefont
  {Wernsdorfer}}]{Ishikawa2005}%
  \BibitemOpen
  \bibfield  {author} {\bibinfo {author} {\bibfnamefont {N.}~\bibnamefont
  {Ishikawa}}, \bibinfo {author} {\bibfnamefont {M.}~\bibnamefont {Sugita}}, \
  and\ \bibinfo {author} {\bibfnamefont {W.}~\bibnamefont {Wernsdorfer}},\
  }\href {\doibase 10.1021/ja0428661} {\bibfield  {journal} {\bibinfo
  {journal} {Journal of the American Chemical Society}\ }\textbf {\bibinfo
  {volume} {127}},\ \bibinfo {pages} {3650} (\bibinfo {year} {2005})},\
  \bibinfo {note} {pMID: 15771471}\BibitemShut {NoStop}%
\bibitem [{\citenamefont {Chen}\ \emph {et~al.}(2017)\citenamefont {Chen},
  \citenamefont {Liu}, \citenamefont {Wernsdorfer}, \citenamefont {Liu},
  \citenamefont {Chibotaru}, \citenamefont {Chen},\ and\ \citenamefont
  {Tong}}]{Chen2017}%
  \BibitemOpen
  \bibfield  {author} {\bibinfo {author} {\bibfnamefont {Y.-C.}\ \bibnamefont
  {Chen}}, \bibinfo {author} {\bibfnamefont {J.-L.}\ \bibnamefont {Liu}},
  \bibinfo {author} {\bibfnamefont {W.}~\bibnamefont {Wernsdorfer}}, \bibinfo
  {author} {\bibfnamefont {D.}~\bibnamefont {Liu}}, \bibinfo {author}
  {\bibfnamefont {L.~F.}\ \bibnamefont {Chibotaru}}, \bibinfo {author}
  {\bibfnamefont {X.-M.}\ \bibnamefont {Chen}}, \ and\ \bibinfo {author}
  {\bibfnamefont {M.-L.}\ \bibnamefont {Tong}},\ }\href {\doibase
  10.1002/anie.201701480} {\bibfield  {journal} {\bibinfo  {journal}
  {Angewandte Chemie International Edition}\ }\textbf {\bibinfo {volume}
  {56}},\ \bibinfo {pages} {4996} (\bibinfo {year} {2017})}\BibitemShut
  {NoStop}%
\bibitem [{\citenamefont {Ding}\ \emph {et~al.}(2018)\citenamefont {Ding},
  \citenamefont {Yu}, \citenamefont {Reta}, \citenamefont {Ortu}, \citenamefont
  {Winpenny}, \citenamefont {Zheng},\ and\ \citenamefont {Chilton}}]{Ding2018}%
  \BibitemOpen
  \bibfield  {author} {\bibinfo {author} {\bibfnamefont {Y.-S.}\ \bibnamefont
  {Ding}}, \bibinfo {author} {\bibfnamefont {K.-X.}\ \bibnamefont {Yu}},
  \bibinfo {author} {\bibfnamefont {D.}~\bibnamefont {Reta}}, \bibinfo {author}
  {\bibfnamefont {F.}~\bibnamefont {Ortu}}, \bibinfo {author} {\bibfnamefont
  {R.~E.~P.}\ \bibnamefont {Winpenny}}, \bibinfo {author} {\bibfnamefont
  {Y.-Z.}\ \bibnamefont {Zheng}}, \ and\ \bibinfo {author} {\bibfnamefont
  {N.~F.}\ \bibnamefont {Chilton}},\ }\href {\doibase
  10.1038/s41467-018-05587-6} {\bibfield  {journal} {\bibinfo  {journal}
  {Nature Communications}\ }\textbf {\bibinfo {volume} {9}},\ \bibinfo {pages}
  {3134} (\bibinfo {year} {2018})}\BibitemShut {NoStop}%
\bibitem [{\citenamefont {Abragam}\ and\ \citenamefont
  {Bleaney}(2012)}]{Abragam2012}%
  \BibitemOpen
  \bibfield  {author} {\bibinfo {author} {\bibfnamefont {A.}~\bibnamefont
  {Abragam}}\ and\ \bibinfo {author} {\bibfnamefont {B.}~\bibnamefont
  {Bleaney}},\ }\href@noop {} {\emph {\bibinfo {title} {Electron Paramagnetic
  Resonance of Transition Ions}}}\ (\bibinfo  {publisher} {Oxford University
  Press},\ \bibinfo {year} {2012})\BibitemShut {NoStop}%
\bibitem [{\citenamefont {Jarmola}\ \emph {et~al.}(2012)\citenamefont
  {Jarmola}, \citenamefont {Acosta}, \citenamefont {Jensen}, \citenamefont
  {Chemerisov},\ and\ \citenamefont {Budker}}]{Jarmola2012}%
  \BibitemOpen
  \bibfield  {author} {\bibinfo {author} {\bibfnamefont {A.}~\bibnamefont
  {Jarmola}}, \bibinfo {author} {\bibfnamefont {V.~M.}\ \bibnamefont {Acosta}},
  \bibinfo {author} {\bibfnamefont {K.}~\bibnamefont {Jensen}}, \bibinfo
  {author} {\bibfnamefont {S.}~\bibnamefont {Chemerisov}}, \ and\ \bibinfo
  {author} {\bibfnamefont {D.}~\bibnamefont {Budker}},\ }\href {\doibase
  10.1103/PhysRevLett.108.197601} {\bibfield  {journal} {\bibinfo  {journal}
  {Phys. Rev. Lett.}\ }\textbf {\bibinfo {volume} {108}},\ \bibinfo {pages}
  {197601} (\bibinfo {year} {2012})}\BibitemShut {NoStop}%
\bibitem [{\citenamefont {Escalera-Moreno}\ \emph
  {et~al.}(2018{\natexlab{b}})\citenamefont {Escalera-Moreno}, \citenamefont
  {Baldoví}, \citenamefont {Gaita-Ariño},\ and\ \citenamefont
  {Coronado}}]{Escalera2018}%
  \BibitemOpen
  \bibfield  {author} {\bibinfo {author} {\bibfnamefont {L.}~\bibnamefont
  {Escalera-Moreno}}, \bibinfo {author} {\bibfnamefont {J.~J.}\ \bibnamefont
  {Baldoví}}, \bibinfo {author} {\bibfnamefont {A.}~\bibnamefont
  {Gaita-Ariño}}, \ and\ \bibinfo {author} {\bibfnamefont {E.}~\bibnamefont
  {Coronado}},\ }\href {\doibase 10.1039/C7SC05464E} {\bibfield  {journal}
  {\bibinfo  {journal} {Chem. Sci.}\ }\textbf {\bibinfo {volume} {9}},\
  \bibinfo {pages} {3265} (\bibinfo {year} {2018}{\natexlab{b}})}\BibitemShut
  {NoStop}%
\bibitem [{\citenamefont {Donati}\ \emph {et~al.}(2020)\citenamefont {Donati},
  \citenamefont {Rusponi}, \citenamefont {Stepanow}, \citenamefont
  {Persichetti}, \citenamefont {Singha}, \citenamefont {Juraschek},
  \citenamefont {W\"ackerlin}, \citenamefont {Baltic}, \citenamefont {Pivetta},
  \citenamefont {Diller}, \citenamefont {Nistor}, \citenamefont {Dreiser},
  \citenamefont {Kummer}, \citenamefont {Velez-Fort}, \citenamefont {Spaldin},
  \citenamefont {Brune},\ and\ \citenamefont {Gambardella}}]{Donati2020}%
  \BibitemOpen
  \bibfield  {author} {\bibinfo {author} {\bibfnamefont {F.}~\bibnamefont
  {Donati}}, \bibinfo {author} {\bibfnamefont {S.}~\bibnamefont {Rusponi}},
  \bibinfo {author} {\bibfnamefont {S.}~\bibnamefont {Stepanow}}, \bibinfo
  {author} {\bibfnamefont {L.}~\bibnamefont {Persichetti}}, \bibinfo {author}
  {\bibfnamefont {A.}~\bibnamefont {Singha}}, \bibinfo {author} {\bibfnamefont
  {D.~M.}\ \bibnamefont {Juraschek}}, \bibinfo {author} {\bibfnamefont
  {C.}~\bibnamefont {W\"ackerlin}}, \bibinfo {author} {\bibfnamefont
  {R.}~\bibnamefont {Baltic}}, \bibinfo {author} {\bibfnamefont
  {M.}~\bibnamefont {Pivetta}}, \bibinfo {author} {\bibfnamefont
  {K.}~\bibnamefont {Diller}}, \bibinfo {author} {\bibfnamefont
  {C.}~\bibnamefont {Nistor}}, \bibinfo {author} {\bibfnamefont
  {J.}~\bibnamefont {Dreiser}}, \bibinfo {author} {\bibfnamefont
  {K.}~\bibnamefont {Kummer}}, \bibinfo {author} {\bibfnamefont
  {E.}~\bibnamefont {Velez-Fort}}, \bibinfo {author} {\bibfnamefont {N.~A.}\
  \bibnamefont {Spaldin}}, \bibinfo {author} {\bibfnamefont {H.}~\bibnamefont
  {Brune}}, \ and\ \bibinfo {author} {\bibfnamefont {P.}~\bibnamefont
  {Gambardella}},\ }\href {\doibase 10.1103/PhysRevLett.124.077204} {\bibfield
  {journal} {\bibinfo  {journal} {Phys. Rev. Lett.}\ }\textbf {\bibinfo
  {volume} {124}},\ \bibinfo {pages} {077204} (\bibinfo {year}
  {2020})}\BibitemShut {NoStop}%
\bibitem [{\citenamefont {Watanabe}\ \emph {et~al.}(2011)\citenamefont
  {Watanabe}, \citenamefont {Yamashita}, \citenamefont {Nakano}, \citenamefont
  {Yamamura},\ and\ \citenamefont {Kajiwara}}]{Watanabe2011}%
  \BibitemOpen
  \bibfield  {author} {\bibinfo {author} {\bibfnamefont {A.}~\bibnamefont
  {Watanabe}}, \bibinfo {author} {\bibfnamefont {A.}~\bibnamefont {Yamashita}},
  \bibinfo {author} {\bibfnamefont {M.}~\bibnamefont {Nakano}}, \bibinfo
  {author} {\bibfnamefont {T.}~\bibnamefont {Yamamura}}, \ and\ \bibinfo
  {author} {\bibfnamefont {T.}~\bibnamefont {Kajiwara}},\ }\href {\doibase
  10.1002/chem.201003538} {\bibfield  {journal} {\bibinfo  {journal} {Chemistry
  – A European Journal}\ }\textbf {\bibinfo {volume} {17}},\ \bibinfo {pages}
  {7428} (\bibinfo {year} {2011})}\BibitemShut {NoStop}%
\bibitem [{\citenamefont {Vallejo}\ \emph {et~al.}(2012)\citenamefont
  {Vallejo}, \citenamefont {Castro}, \citenamefont {Ruiz-García},
  \citenamefont {Cano}, \citenamefont {Julve}, \citenamefont {Lloret},
  \citenamefont {De~Munno}, \citenamefont {Wernsdorfer},\ and\ \citenamefont
  {Pardo}}]{Vallejo2012}%
  \BibitemOpen
  \bibfield  {author} {\bibinfo {author} {\bibfnamefont {J.}~\bibnamefont
  {Vallejo}}, \bibinfo {author} {\bibfnamefont {I.}~\bibnamefont {Castro}},
  \bibinfo {author} {\bibfnamefont {R.}~\bibnamefont {Ruiz-García}}, \bibinfo
  {author} {\bibfnamefont {J.}~\bibnamefont {Cano}}, \bibinfo {author}
  {\bibfnamefont {M.}~\bibnamefont {Julve}}, \bibinfo {author} {\bibfnamefont
  {F.}~\bibnamefont {Lloret}}, \bibinfo {author} {\bibfnamefont
  {G.}~\bibnamefont {De~Munno}}, \bibinfo {author} {\bibfnamefont
  {W.}~\bibnamefont {Wernsdorfer}}, \ and\ \bibinfo {author} {\bibfnamefont
  {E.}~\bibnamefont {Pardo}},\ }\href {\doibase 10.1021/ja3075314} {\bibfield
  {journal} {\bibinfo  {journal} {Journal of the American Chemical Society}\
  }\textbf {\bibinfo {volume} {134}},\ \bibinfo {pages} {15704} (\bibinfo
  {year} {2012})},\ \bibinfo {note} {pMID: 22963111},\ \Eprint
  {http://arxiv.org/abs/https://doi.org/10.1021/ja3075314}
  {https://doi.org/10.1021/ja3075314} \BibitemShut {NoStop}%
\bibitem [{\citenamefont {Gomez-Coca}\ \emph {et~al.}(2013)\citenamefont
  {Gomez-Coca}, \citenamefont {Cremades}, \citenamefont {Aliaga-Alcalde},\ and\
  \citenamefont {Ruiz}}]{Coca2013}%
  \BibitemOpen
  \bibfield  {author} {\bibinfo {author} {\bibfnamefont {S.}~\bibnamefont
  {Gomez-Coca}}, \bibinfo {author} {\bibfnamefont {E.}~\bibnamefont
  {Cremades}}, \bibinfo {author} {\bibfnamefont {N.}~\bibnamefont
  {Aliaga-Alcalde}}, \ and\ \bibinfo {author} {\bibfnamefont {E.}~\bibnamefont
  {Ruiz}},\ }\href {\doibase 10.1021/ja4015138} {\bibfield  {journal} {\bibinfo
   {journal} {Journal of the American Chemical Society}\ }\textbf {\bibinfo
  {volume} {135}},\ \bibinfo {pages} {7010} (\bibinfo {year} {2013})},\
  \bibinfo {note} {pMID: 23586965},\ \Eprint
  {http://arxiv.org/abs/https://doi.org/10.1021/ja4015138}
  {https://doi.org/10.1021/ja4015138} \BibitemShut {NoStop}%
\bibitem [{\citenamefont {Zhu}\ \emph {et~al.}(2013)\citenamefont {Zhu},
  \citenamefont {Cui}, \citenamefont {Zhang}, \citenamefont {Jia},
  \citenamefont {Guo}, \citenamefont {Gao}, \citenamefont {Qian}, \citenamefont
  {Jiang}, \citenamefont {Wang}, \citenamefont {Wang},\ and\ \citenamefont
  {Gao}}]{Zhu2013}%
  \BibitemOpen
  \bibfield  {author} {\bibinfo {author} {\bibfnamefont {Y.-Y.}\ \bibnamefont
  {Zhu}}, \bibinfo {author} {\bibfnamefont {C.}~\bibnamefont {Cui}}, \bibinfo
  {author} {\bibfnamefont {Y.-Q.}\ \bibnamefont {Zhang}}, \bibinfo {author}
  {\bibfnamefont {J.-H.}\ \bibnamefont {Jia}}, \bibinfo {author} {\bibfnamefont
  {X.}~\bibnamefont {Guo}}, \bibinfo {author} {\bibfnamefont {C.}~\bibnamefont
  {Gao}}, \bibinfo {author} {\bibfnamefont {K.}~\bibnamefont {Qian}}, \bibinfo
  {author} {\bibfnamefont {S.-D.}\ \bibnamefont {Jiang}}, \bibinfo {author}
  {\bibfnamefont {B.-W.}\ \bibnamefont {Wang}}, \bibinfo {author}
  {\bibfnamefont {Z.-M.}\ \bibnamefont {Wang}}, \ and\ \bibinfo {author}
  {\bibfnamefont {S.}~\bibnamefont {Gao}},\ }\href {\doibase
  10.1039/C3SC21893G} {\bibfield  {journal} {\bibinfo  {journal} {Chem. Sci.}\
  }\textbf {\bibinfo {volume} {4}},\ \bibinfo {pages} {1802} (\bibinfo {year}
  {2013})}\BibitemShut {NoStop}%
\bibitem [{\citenamefont {Fataftah}\ \emph {et~al.}(2014)\citenamefont
  {Fataftah}, \citenamefont {Zadrozny}, \citenamefont {Rogers},\ and\
  \citenamefont {Freedman}}]{Fataftah2014}%
  \BibitemOpen
  \bibfield  {author} {\bibinfo {author} {\bibfnamefont {M.~S.}\ \bibnamefont
  {Fataftah}}, \bibinfo {author} {\bibfnamefont {J.~M.}\ \bibnamefont
  {Zadrozny}}, \bibinfo {author} {\bibfnamefont {D.~M.}\ \bibnamefont
  {Rogers}}, \ and\ \bibinfo {author} {\bibfnamefont {D.~E.}\ \bibnamefont
  {Freedman}},\ }\href {\doibase 10.1021/ic501906z} {\bibfield  {journal}
  {\bibinfo  {journal} {Inorganic Chemistry}\ }\textbf {\bibinfo {volume}
  {53}},\ \bibinfo {pages} {10716} (\bibinfo {year} {2014})},\ \bibinfo {note}
  {pMID: 25198379}\BibitemShut {NoStop}%
\bibitem [{\citenamefont {Pedersen}\ \emph {et~al.}(2015)\citenamefont
  {Pedersen}, \citenamefont {Dreiser}, \citenamefont {Weihe}, \citenamefont
  {Sibille}, \citenamefont {Johannesen}, \citenamefont {Sørensen},
  \citenamefont {Nielsen}, \citenamefont {Sigrist}, \citenamefont {Mutka},
  \citenamefont {Rols}, \citenamefont {Bendix},\ and\ \citenamefont
  {Piligkos}}]{Pedersen2015}%
  \BibitemOpen
  \bibfield  {author} {\bibinfo {author} {\bibfnamefont {K.~S.}\ \bibnamefont
  {Pedersen}}, \bibinfo {author} {\bibfnamefont {J.}~\bibnamefont {Dreiser}},
  \bibinfo {author} {\bibfnamefont {H.}~\bibnamefont {Weihe}}, \bibinfo
  {author} {\bibfnamefont {R.}~\bibnamefont {Sibille}}, \bibinfo {author}
  {\bibfnamefont {H.~V.}\ \bibnamefont {Johannesen}}, \bibinfo {author}
  {\bibfnamefont {M.~A.}\ \bibnamefont {Sørensen}}, \bibinfo {author}
  {\bibfnamefont {B.~E.}\ \bibnamefont {Nielsen}}, \bibinfo {author}
  {\bibfnamefont {M.}~\bibnamefont {Sigrist}}, \bibinfo {author} {\bibfnamefont
  {H.}~\bibnamefont {Mutka}}, \bibinfo {author} {\bibfnamefont
  {S.}~\bibnamefont {Rols}}, \bibinfo {author} {\bibfnamefont {J.}~\bibnamefont
  {Bendix}}, \ and\ \bibinfo {author} {\bibfnamefont {S.}~\bibnamefont
  {Piligkos}},\ }\href {\doibase 10.1021/acs.inorgchem.5b01209} {\bibfield
  {journal} {\bibinfo  {journal} {Inorganic Chemistry}\ }\textbf {\bibinfo
  {volume} {54}},\ \bibinfo {pages} {7600} (\bibinfo {year} {2015})},\ \bibinfo
  {note} {pMID: 26201004}\BibitemShut {NoStop}%
\bibitem [{\citenamefont {Novikov}\ \emph {et~al.}(2015)\citenamefont
  {Novikov}, \citenamefont {Pavlov}, \citenamefont {Nelyubina}, \citenamefont
  {Boulon}, \citenamefont {Varzatskii}, \citenamefont {Voloshin},\ and\
  \citenamefont {Winpenny}}]{Novikov2015}%
  \BibitemOpen
  \bibfield  {author} {\bibinfo {author} {\bibfnamefont {V.~V.}\ \bibnamefont
  {Novikov}}, \bibinfo {author} {\bibfnamefont {A.~A.}\ \bibnamefont {Pavlov}},
  \bibinfo {author} {\bibfnamefont {Y.~V.}\ \bibnamefont {Nelyubina}}, \bibinfo
  {author} {\bibfnamefont {M.-E.}\ \bibnamefont {Boulon}}, \bibinfo {author}
  {\bibfnamefont {O.~A.}\ \bibnamefont {Varzatskii}}, \bibinfo {author}
  {\bibfnamefont {Y.~Z.}\ \bibnamefont {Voloshin}}, \ and\ \bibinfo {author}
  {\bibfnamefont {R.~E.}\ \bibnamefont {Winpenny}},\ }\href {\doibase
  10.1021/jacs.5b05739} {\bibfield  {journal} {\bibinfo  {journal} {Journal of
  the American Chemical Society}\ }\textbf {\bibinfo {volume} {137}},\ \bibinfo
  {pages} {9792} (\bibinfo {year} {2015})},\ \bibinfo {note} {pMID:
  26199996}\BibitemShut {NoStop}%
\bibitem [{\citenamefont {Rechkemmer}\ \emph {et~al.}(2016)\citenamefont
  {Rechkemmer}, \citenamefont {Breitgoff}, \citenamefont {van~der Meer},
  \citenamefont {Atanasov}, \citenamefont {Hakl}, \citenamefont {Orlita},
  \citenamefont {Neugebauer}, \citenamefont {Neese}, \citenamefont {Sarkar},\
  and\ \citenamefont {van Slageren}}]{Rechkemmer2016}%
  \BibitemOpen
  \bibfield  {author} {\bibinfo {author} {\bibfnamefont {Y.}~\bibnamefont
  {Rechkemmer}}, \bibinfo {author} {\bibfnamefont {F.~D.}\ \bibnamefont
  {Breitgoff}}, \bibinfo {author} {\bibfnamefont {M.}~\bibnamefont {van~der
  Meer}}, \bibinfo {author} {\bibfnamefont {M.}~\bibnamefont {Atanasov}},
  \bibinfo {author} {\bibfnamefont {M.}~\bibnamefont {Hakl}}, \bibinfo {author}
  {\bibfnamefont {M.}~\bibnamefont {Orlita}}, \bibinfo {author} {\bibfnamefont
  {P.}~\bibnamefont {Neugebauer}}, \bibinfo {author} {\bibfnamefont
  {F.}~\bibnamefont {Neese}}, \bibinfo {author} {\bibfnamefont
  {B.}~\bibnamefont {Sarkar}}, \ and\ \bibinfo {author} {\bibfnamefont
  {J.}~\bibnamefont {van Slageren}},\ }\href {\doibase 10.1038/ncomms10467}
  {\bibfield  {journal} {\bibinfo  {journal} {Nature Communications}\ }\textbf
  {\bibinfo {volume} {7}},\ \bibinfo {pages} {10467} (\bibinfo {year}
  {2016})}\BibitemShut {NoStop}%
\bibitem [{\citenamefont {Wang}\ \emph {et~al.}(2019)\citenamefont {Wang},
  \citenamefont {Ruan}, \citenamefont {Li}, \citenamefont {Chen}, \citenamefont
  {Huang}, \citenamefont {Liu}, \citenamefont {Reta}, \citenamefont {Chilton},
  \citenamefont {Wang},\ and\ \citenamefont {Tong}}]{Wang2019}%
  \BibitemOpen
  \bibfield  {author} {\bibinfo {author} {\bibfnamefont {J.}~\bibnamefont
  {Wang}}, \bibinfo {author} {\bibfnamefont {Z.-Y.}\ \bibnamefont {Ruan}},
  \bibinfo {author} {\bibfnamefont {Q.-W.}\ \bibnamefont {Li}}, \bibinfo
  {author} {\bibfnamefont {Y.-C.}\ \bibnamefont {Chen}}, \bibinfo {author}
  {\bibfnamefont {G.-Z.}\ \bibnamefont {Huang}}, \bibinfo {author}
  {\bibfnamefont {J.-L.}\ \bibnamefont {Liu}}, \bibinfo {author} {\bibfnamefont
  {D.}~\bibnamefont {Reta}}, \bibinfo {author} {\bibfnamefont {N.~F.}\
  \bibnamefont {Chilton}}, \bibinfo {author} {\bibfnamefont {Z.-X.}\
  \bibnamefont {Wang}}, \ and\ \bibinfo {author} {\bibfnamefont {M.-L.}\
  \bibnamefont {Tong}},\ }\href {\doibase 10.1039/C8DT04814B} {\bibfield
  {journal} {\bibinfo  {journal} {Dalton Trans.}\ }\textbf {\bibinfo {volume}
  {48}},\ \bibinfo {pages} {1686} (\bibinfo {year} {2019})}\BibitemShut
  {NoStop}%
\bibitem [{\citenamefont {Kobayashi}\ \emph {et~al.}(2019)\citenamefont
  {Kobayashi}, \citenamefont {Ohtani}, \citenamefont {Nakamura}, \citenamefont
  {Lindoy},\ and\ \citenamefont {Hayami}}]{Kobayashi2019}%
  \BibitemOpen
  \bibfield  {author} {\bibinfo {author} {\bibfnamefont {F.}~\bibnamefont
  {Kobayashi}}, \bibinfo {author} {\bibfnamefont {R.}~\bibnamefont {Ohtani}},
  \bibinfo {author} {\bibfnamefont {M.}~\bibnamefont {Nakamura}}, \bibinfo
  {author} {\bibfnamefont {L.~F.}\ \bibnamefont {Lindoy}}, \ and\ \bibinfo
  {author} {\bibfnamefont {S.}~\bibnamefont {Hayami}},\ }\href {\doibase
  10.1021/acs.inorgchem.9b00543} {\bibfield  {journal} {\bibinfo  {journal}
  {Inorganic Chemistry}\ }\textbf {\bibinfo {volume} {58}},\ \bibinfo {pages}
  {7409} (\bibinfo {year} {2019})},\ \bibinfo {note} {pMID:
  31117627}\BibitemShut {NoStop}%
\bibitem [{\citenamefont {Breuer}(2007)}]{Breuer2007}%
  \BibitemOpen
  \bibfield  {author} {\bibinfo {author} {\bibfnamefont {H.-P.}\ \bibnamefont
  {Breuer}},\ }\href@noop {} {\emph {\bibinfo {title} {The Theory of Open
  Quantum Systems}}}\ (\bibinfo  {publisher} {Oxford University Press},\
  \bibinfo {year} {2007})\BibitemShut {NoStop}%
\bibitem [{\citenamefont {Stefanucci}\ and\ \citenamefont {van
  Leeuwe}(2013)}]{Stefanucci2013}%
  \BibitemOpen
  \bibfield  {author} {\bibinfo {author} {\bibfnamefont {G.}~\bibnamefont
  {Stefanucci}}\ and\ \bibinfo {author} {\bibfnamefont {R.}~\bibnamefont {van
  Leeuwe}},\ }\href@noop {} {\emph {\bibinfo {title} {Nonequilibrium Many-Body
  Theory of Quantum Systems: A Modern Introduction}}},\ \bibinfo {edition}
  {1st}\ ed.\ (\bibinfo  {publisher} {Cambridge University Press},\ \bibinfo
  {year} {2013})\BibitemShut {NoStop}%
\bibitem [{\citenamefont {Xu}\ \emph {et~al.}(2008)\citenamefont {Xu},
  \citenamefont {Wang}, \citenamefont {Duan}, \citenamefont {Gu},\ and\
  \citenamefont {Li}}]{Xu2008}%
  \BibitemOpen
  \bibfield  {author} {\bibinfo {author} {\bibfnamefont {Y.}~\bibnamefont
  {Xu}}, \bibinfo {author} {\bibfnamefont {J.-S.}\ \bibnamefont {Wang}},
  \bibinfo {author} {\bibfnamefont {W.}~\bibnamefont {Duan}}, \bibinfo {author}
  {\bibfnamefont {B.-L.}\ \bibnamefont {Gu}}, \ and\ \bibinfo {author}
  {\bibfnamefont {B.}~\bibnamefont {Li}},\ }\href {\doibase
  10.1103/PhysRevB.78.224303} {\bibfield  {journal} {\bibinfo  {journal} {Phys.
  Rev. B}\ }\textbf {\bibinfo {volume} {78}},\ \bibinfo {pages} {224303}
  (\bibinfo {year} {2008})}\BibitemShut {NoStop}%
\bibitem [{\citenamefont {Neese}(2018)}]{Neese2018}%
  \BibitemOpen
  \bibfield  {author} {\bibinfo {author} {\bibfnamefont {F.}~\bibnamefont
  {Neese}},\ }\href {\doibase 10.1002/wcms.1327} {\bibfield  {journal}
  {\bibinfo  {journal} {WIREs Computational Molecular Science}\ }\textbf
  {\bibinfo {volume} {8}},\ \bibinfo {pages} {e1327} (\bibinfo {year}
  {2018})}\BibitemShut {NoStop}%
\bibitem [{\citenamefont {Perdew}\ \emph {et~al.}(1996)\citenamefont {Perdew},
  \citenamefont {Burke},\ and\ \citenamefont {Wang}}]{PBE}%
  \BibitemOpen
  \bibfield  {author} {\bibinfo {author} {\bibfnamefont {J.~P.}\ \bibnamefont
  {Perdew}}, \bibinfo {author} {\bibfnamefont {K.}~\bibnamefont {Burke}}, \
  and\ \bibinfo {author} {\bibfnamefont {Y.}~\bibnamefont {Wang}},\ }\href
  {\doibase 10.1103/PhysRevB.54.16533} {\bibfield  {journal} {\bibinfo
  {journal} {Phys. Rev. B}\ }\textbf {\bibinfo {volume} {54}},\ \bibinfo
  {pages} {16533} (\bibinfo {year} {1996})}\BibitemShut {NoStop}%
\bibitem [{\citenamefont {Atanasov}\ \emph {et~al.}(2015)\citenamefont
  {Atanasov}, \citenamefont {Aravena}, \citenamefont {Suturina}, \citenamefont
  {Bill}, \citenamefont {Maganas},\ and\ \citenamefont {Neese}}]{Atanasov2015}%
  \BibitemOpen
  \bibfield  {author} {\bibinfo {author} {\bibfnamefont {M.}~\bibnamefont
  {Atanasov}}, \bibinfo {author} {\bibfnamefont {D.}~\bibnamefont {Aravena}},
  \bibinfo {author} {\bibfnamefont {E.}~\bibnamefont {Suturina}}, \bibinfo
  {author} {\bibfnamefont {E.}~\bibnamefont {Bill}}, \bibinfo {author}
  {\bibfnamefont {D.}~\bibnamefont {Maganas}}, \ and\ \bibinfo {author}
  {\bibfnamefont {F.}~\bibnamefont {Neese}},\ }\href {\doibase
  https://doi.org/10.1016/j.ccr.2014.10.015} {\bibfield  {journal} {\bibinfo
  {journal} {Coordination Chemistry Reviews}\ }\textbf {\bibinfo {volume}
  {289-290}},\ \bibinfo {pages} {177 } (\bibinfo {year} {2015})}\BibitemShut
  {NoStop}%
\bibitem [{\citenamefont {Lyo}(1972)}]{Lyo1972}%
  \BibitemOpen
  \bibfield  {author} {\bibinfo {author} {\bibfnamefont {S.~K.}\ \bibnamefont
  {Lyo}},\ }\href {\doibase 10.1103/PhysRevB.5.795} {\bibfield  {journal}
  {\bibinfo  {journal} {Phys. Rev. B}\ }\textbf {\bibinfo {volume} {5}},\
  \bibinfo {pages} {795} (\bibinfo {year} {1972})}\BibitemShut {NoStop}%
\end{thebibliography}%
\end{document}